\documentclass[12pt]{article}
\pdfoutput=1 
\usepackage[utf8]{inputenc}
\usepackage[T1]{fontenc}

\usepackage{setspace} % must come before hyperref
\usepackage[vmargin = 1.5in, hmargin =1.5in]{geometry}

\usepackage[usenames,dvipsnames,svgnames]{xcolor}
\usepackage{amsmath, amssymb, amsfonts, graphicx, tikz,pdflscape, mathtools, amsthm, upgreek, bm,pgfplots,csvsimple,multirow, multicol, booktabs}
\usepackage{verbatim}
\usepackage{natbib}
\usepackage{accents}
\newcommand{\ubar}[1]{\underaccent{\bar}{#1}}
\usepackage{needspace}
\def\citeapos#1{\citeauthor{#1}'s (\citeyear{#1})}
\usepackage{comment}
\usepackage[inline]{enumitem}
\usepackage{tocvsec2}
\usepackage{threeparttable}
\usetikzlibrary{calc, cd}
\setlist{noitemsep, topsep=0pt}
\usepackage[skip = 10pt]{caption}
\allowdisplaybreaks[1]
\allowdisplaybreaks[1]

\definecolor{DarkBlue}{RGB}{0, 114,178}
\definecolor{Blue}{RGB}{43,147,206}
\definecolor{LightBlue}{RGB}{86,180,233}

\definecolor{DarkOrange}{RGB}{213,94,0}
\definecolor{Orange}{RGB}{221,126,0}
\definecolor{LightOrange}{RGB}{230,159,0}

\definecolor{Green}{RGB}{0,158,115} % graphics color 3
\definecolor{Purple}{RGB}{204,121,167} % graphics color 4
\definecolor{Gray}{RGB}{0,0,150} % de-emphasized color
\definecolor{Ivory}{RGB}{255,255,240} % background color

\definecolor{TextColor1}{RGB}{0, 114,178}
\definecolor{TextColor2}{RGB}{213,94,0}
\definecolor{PlotColor1}{RGB}{43,147,206}
\definecolor{PlotColor2}{RGB}{221,126,0}
\definecolor{PlotColor3}{RGB}{0,158,115}
\definecolor{BackgroundColor}{RGB}{255,255,240}
\definecolor{GmailBlue}{RGB}{42, 93, 176} % for links
\usepackage{hyperref}
\hypersetup{
		colorlinks=true,
		citecolor= GmailBlue,
		linkcolor=GmailBlue,
		urlcolor = GmailBlue}
\newcommand{\bibtexorder}[1]{}

\usepackage{pgfplots}
\usepgfplotslibrary{groupplots}
\newcommand{\plotgap}{2.5 cm} %% for side by side plots
\usepackage{tikz}
\usetikzlibrary{matrix,calc,shapes,arrows.meta,positioning}
\pgfplotsset{width = \textwidth/2}
\tikzstyle{hollow}=[circle,draw,inner sep=1.5]
\tikzstyle{solid}=[circle,draw,inner sep=1.5,fill=black]
\pgfplotsset{compat = newest}

\usepackage[capitalize,noabbrev]{cleveref}

\newtheoremstyle{breakital}% name
{}%         Space above, empty = `usual value'
{}%         Space below
{\itshape}% Body font
{}%         Indent amount (empty = no indent, \parindent = para indent)
{\bfseries}% Thm head font
{}%        Punctuation after thm head
{\newline}% Space after thm head: \newline = linebreak
{}%         Thm head spec

\theoremstyle{breakital}
\newtheorem{thm}{Theorem}
\newtheorem*{theorem*}{Theorem}
\newtheorem*{cor*}{Corollary}

\newtheorem{prop}{Proposition}
\newtheorem{lem}{Lemma}

\crefname{prop}{Proposition}{Propositions}
\crefname{thm}{Theorem}{Theorems}
\crefname{lem}{Lemma}{Lemmas}

\newtheoremstyle{break}% name
{}%         Space above, empty = `usual value'
{}%         Space below
{}% Body font
{}%         Indent amount (empty = no indent, \parindent = para indent)
{\bfseries}% Thm head font
{}%        Punctuation after thm head
{\newline}% Space after thm head: \newline = linebreak
{}%         Thm head spec

\theoremstyle{break}
\newtheorem{exmp}{Example}
\newtheorem*{as*}{Assumptions}
\crefname{as}{Assumption}{Assumptions}

\theoremstyle{definition}

\newtheorem*{rem*}{Remark}

\numberwithin{lem2}{section}
\crefname{lem2}{Lemma}{Lemmas}

\def\a{\alpha}
\def\b{\beta}
\def\g{\gamma}
\def\d{\delta}
\def\e{\varepsilon}

\def\h{\eta}
\def\th{\theta}

\def\k{\kappa}
\def\l{\lambda}

\def\s{\sigma}

\label{key}

\def\D{\Delta}

\def\S{\Sigma}

\def\R{\mathbf{R}}

\def\P{\mathbf{P}}

\DeclareMathOperator{\expec}{\mathbf{E}}
\DeclareMathOperator{\E}{\mathbf{E}}
\DeclareMathOperator{\supp}{supp} % support

\DeclareMathOperator{\diag}{diag}

\DeclareMathOperator{\Cov}{cov} % covariance
\DeclareMathOperator{\cov}{cov} % covariance
\DeclareMathOperator{\reg}{reg} % covariance
\DeclareMathOperator{\var}{var}
\DeclareMathOperator{\signal}{signal}
\DeclareMathOperator{\score}{score}
\DeclareMathOperator{\screen}{screen}

\DeclareMathOperator{\BR}{BR}

\newcommand{\Brac}[1]{\left[ #1 \right]}

\title{Scoring Strategic Agents%
}

\author{Ian Ball\thanks{%
		MIT Department of Economics, ianball@mit.edu.
		This paper was previously circulated with the title ``Incentive-Compatible Prediction.'' An earlier version of this paper appeared as the first chapter of my dissertation at Yale. For continual guidance, I thank my advisor Dirk Bergemann and my committee members Larry Samuelson and Johannes H\"{o}rner.
		For helpful comments, I thank
		Nageeb Ali,
		Isaiah Andrews,
		Simon Board,
		Alessandro Bonatti,
		Tilman B\"{o}rgers,
		Hector Chade,
		Gonzalo Cisternas, 
		Joyee Deb,
		Jeff Ely,
		Jos\'{e}-Antonio Esp\'{i}n-S\'{a}nchez,
		Mira Frick,
		Marina Halac,
		Rick Harbaugh,
		Tibor Heumann,
		Ryota Iijima,
		Navin Kartik,
		Deniz Kattwinkel,
		Jan Knoepfle,
		Soonwoo Kwon,
		Patrick Lahr,
		Stephan Lauermann,
		Ro'ee Levy,
		Xiangliang Li,
		Heng Liu,
		Niccol\`{o} Lomys,
		Erik Madsen,
		George Mailath,
		Chiara Margaria,
		Idione Meneghel,
		Meg Meyer,
		Konrad Mierendorff,
		Weicheng Min,
		Stephen Morris,
		Xiaosheng Mu,
		Miquel Oliu-Barton, 
		Juan Ortner,
		Yujie Qian,
		Tim Roughgarden,
		Anna Rubinchik,
		Grant Schoenebeck,
		Eran Shmaya,
		Nicholas Snashall-Woodhams,
		Suk Joon Son,
		Rani Spiegler,
		Philipp Strack,
		Juuso V\"{a}lim\"{a}ki,
		Nisheeth Vishnoi,
		Allen Vong,
		Conor Walsh,
		Xinyang Wang,
		and
		Weijie Zhong.
	}
}

\date{15 May 2024}

\begin{document}

\maketitle

\begin{abstract} 
I introduce a model of predictive scoring. A receiver wants to predict a sender's quality. An intermediary observes multiple features of the sender and aggregates them into a score. Based on the score, the receiver makes a decision. The sender prefers ``higher'' decisions, and she can distort each feature at a privately known cost. I characterize the scoring rule that maximizes decision accuracy. This rule underweights some features to deter sender distortion, and overweights other features so that the score is correct on average. The receiver prefers this scoring rule to full disclosure because it mitigates his commitment problem. 
\end{abstract}
 

\begin{comment}

%tableofcontents
\setcounter{tocdepth}{4}
\newpage
{
   \singlespacing
  \hypersetup{linkcolor=black}
  \tableofcontents 
}
\end{comment}

\newpage

\section{Introduction} \label{sec:introduction}

%\subsection{Motivation and results} \label{sec:motivation}

% Scoring - define and motivate
As data sources proliferate, predictive scores are increasingly used to guide important decisions. Banks use credit scores to set the terms of loans; judges use defendant risk scores to set bail; and online platforms score sellers, businesses, and job-seekers. We now live in a ``scored society'' \citep{CitronPasquale2014}. These scores have a common structure: an intermediary gathers data about an agent from different sources and then aggregates the agent’s features into a score that predicts a quality of interest. For example, a FICO credit score aggregates features such as credit utilization rate and length of credit history in order to predict a consumer's creditworthiness.%
%\footnote{According to \href{https://ficoscore.com}{FICO}, its scores are used in 90\% of U.S. lending decisions.} 

% Challenge - strategic behavior; then research question
Predictive scoring is not only a statistical problem. Strategic manipulation presents an additional challenge, commonly known as Goodhart's law \citep{Goodhart1975}. An agent who understands the scoring rule can distort her features to improve her score, without changing her quality, thus undermining the accuracy of the score. For example, a consumer can spread her spending across multiple credit cards to lower her credit utilization rate, without reducing her risk of default. ``The scoring models may not be telling us the same thing that they have historically,''  according to Mark Zandi, chief economist at Moody's, ``because people are so focused on their scores and working hard to get them up.''%
\footnote{``How More Americans Are Getting a Perfect Credit Score'' (\href{https://www.bloomberg.com/news/features/2017-08-14/obsessives-have-cracked-the-perfect-fico-credit-score-of-850?embedded-checkout=true}{\textit{Bloomberg}, 2017}).}
Across different contexts, whenever scores influence high-stakes decisions, these scores tend to be manipulated.%
\footnote{For instance, hospitals admit healthy patients to improve their scorecards, and law schools hire their own graduates to boost their US News rankings; see \cite{EdererHoldenMeyer2018}.}
In the presence of such strategic behavior, what scoring rule induces the most accurate decisions? 

% To answer research question, build a model of scoring
To answer this question, I build a model of predictive scoring. There are three players: sender (she), intermediary, and receiver (he). The sender is the agent being scored. The receiver wants to predict the \emph{quality} of the sender. The intermediary observes multiple manipulable \emph{features} of the sender, and commits to a rule for aggregating these features into a score (from an unrestricted set). The receiver sees this score and makes a decision. He wants his decision to match the sender’s quality. The sender prefers higher decisions, and she can distort each of her features at a cost. For each feature, the sender has two kinds of private information. The \emph{intrinsic level} is the value the feature would take if the sender did not distort it. The \emph{distortion ability} parameterizes the sender's cost of distorting that feature. The sender's quality is correlated with the intrinsic levels of the features. 
% Intermediary's problem - different from statistical problem 
%The intermediary has the same preferences as the receiver. Specifically,

The intermediary designs the scoring rule to minimize the mean squared error between the receiver's decision and the sender's quality. If the sender's features were exogenous, then predicting her quality would be a purely statistical problem. Instead, the intermediary must consider how the scoring rule motivates the sender to distort her features. Formally, each scoring rule induces a different game between the sender and the receiver. 

My main contribution is  introducing and analyzing this model of scoring, which lies between signaling and screening in the level of commitment afforded to the receiver. I analyze all three settings,  in order of increasing commitment: signaling, scoring, and then screening.
% Signaling: it is a benchmark

First, I consider a \emph{signaling} setting without the intermediary. The receiver observes the sender's distorted features, which serve as signals of the sender's quality. This signaling setting gives a lower bound on the intermediary's payoff from optimal scoring since full disclosure is a feasible scoring rule.

%Any equilibrium outcome of this signaling benchmark is achievable under scoring since full disclosure is one feasible scoring rule for the intermediary. 

% an While there is no intermediary in this setting, ththe interpretation interpret this benchmark Formally, there is no intermediary in this setting, but I interpret it as the this setting as a benchm the interpretation This benchmark gives a lower bound on the receiver's utility under optimal scoring: The intermediary can replicate the signaling game by fully disclosing the sender's features.

% Signaling - information loss from heterogeneity
In the signaling game, the sender's distortion can interfere with the receiver's prediction of the sender's quality. I first show that the signaling game has exactly one equilibrium in linear strategies (\cref{res:existence_uniqueness}). In this equilibrium, the sender distorts each feature in proportion to her distortion ability on that feature. If distortion ability is homogeneous in the population, then every sender type distorts her features by the same amount. The receiver anticipates this distortion and subtracts it in order to recover the sender's intrinsic levels. If instead distortion ability is heterogeneous, then each feature confounds the sender's intrinsic level with her distortion ability. The receiver cannot distinguish a sender with a high intrinsic level and low distortion ability from a sender with a lower intrinsic level but higher distortion ability.

% Intermediary mitigates commitment problem 
Next, I return to the main \emph{scoring} setting in which the receiver learns about the sender's quality only through the intermediary's score. While more information always helps a decision-maker acting in isolation, here information coarsening can improve the receiver's predictions by mitigating a commitment problem. To illustrate this problem, suppose that the receiver tried to discourage distortion by making his decision less sensitive to the sender's features. In response, the sender would distort her features less. Her features would then be more informative about her quality, so the receiver would want to make his decision \emph{more} sensitive to them. 

% Sequential rationality requires the receiver's decision to be optimal given the distribution of the sender's features. Hence, the receiver cannot internalize the effect of his strategy on the sender's choice of distortion. But this strategy is not sequentially rational. 

% sharpening value of scoring
I restrict attention to linear scoring rules. First, I characterize the type distributions for which optimal scoring yields strictly more accurate decisions than the signaling equilibrium (\cref{res:scoring_equals_signaling}). My condition holds generically. 
Under a further restriction on the covariance structure---termed uncorrelated errors---I analyze the scoring rule feature-by-feature (\cref{res:underweighting_overweighting}). When the signaling and scoring solutions differ, the optimal scoring rule underweights some features in order to deter sender distortion, and it overweights other features so that the score remains correct on average. A feature is more  underweighted if distortion ability is more heterogeneous or if the intrinsic level is more informative of quality. If the receiver could observe the sender's features ex post, he would choose a different decision, but from the score alone he cannot disentangle the value of each feature. 

% The receiver has fewer feasible deviations. Since the receiver has fewer feasible deviations, his commitment problem is mitigated.or then dampening distortion has the greatest informational benefit. ).  
% In particular, the receiver can commit to decisions that are ex post suboptimal.
% Screening -  underweighting every feature; commitment comparison
Finally, I consider a \emph{screening} setting in which the receiver can commit to his decision as a function of the sender's features. Again, I focus on linear rules. As the receiver's commitment increases---from no commitment (signaling) to information commitment (scoring) to decision commitment (screening)---the information loss from the sender's distortion decreases (\cref{res:reduced_distortion}). With uncorrelated errors, 
each feature weight under the optimal screening rule is smaller than the corresponding weight under the optimal scoring rule (\cref{res:feature_weights}). Therefore, the average distortion costs for the sender are lower under screening than under scoring. 

%With full commitment power, there is no reason to overweight any of the sender's features. 
%Unlike in the scoring setting, the receiver can underweight every feature simultaneously

%While the receiver benefits from additional commitment, the welfare comparison for the sender is ambiguous. With commitment power, the receiver selects the feature weights according to the heterogeneity of the sender's distortion, but the sender's cost depends on the amount of distortion, not its heterogeneity. 

% Extensions
%Finally, I extend the main model. First, I allow for stochastic scoring rules. I give conditions under which the optimal scoring rule is noise-free. Next I consider a different objective for the intermediary. If the intermediary maximizes social welfare rather than information transmission, then adding noise to the scoring rule is optimal because it reduces the sender's inefficient distortion. Indeed, if the objective were to maximize the sender's utility, then the optimal policy would be to provide no information. 

The rest of the paper is organized as follows. \cref{sec:literature} discusses related literature. \cref{sec:model} presents the model of predictive scoring.  \cref{sec:signaling,sec:scoring,sec:screening} study signaling, scoring, and screening, respectively. \cref{sec:extensions} shows that random linear scores cannot outperform the deterministic linear scores considered in the main model. The conclusion is in \cref{sec:conclusion}.  Proofs are in \cref{sec:proofs}.  \cref{sec:beyond_linear} gives an additional result about nonlinear equilibria of the signaling game.

\subsection{Related literature} \label{sec:literature}

The scoring problem that I consider is interesting only in signaling settings with (a) multiple signals/features and (b) sender preferences that violate single-crossing. Previous signaling models have allowed (a) or (b) separately, but not together. The classical multi-dimensional signaling environments  in \cite{Engers1987} and \cite{QuinziiRochet1985} admit fully separating equilibria, so there is no scope for intermediation to improve information transmission. The closest work on signaling without single-crossing considers a single signal, with a
linear--quadratic--Gaussian/elliptical (LQG/E) structure \citep{PrendergastTopel1996,FischerVerrecchia2000, BenabouTirole2006, Gesche2021,FrankelKartik2019}.\footnote{\cite{FrankelKartik2019} also consider a general setting, without the LQE assumption.} The equilibria in these models are not fully separating, but with only one signal, there is no scope for signal \emph{aggregation} by an intermediary.

\begin{comment}
  In the single-signal elliptical environment of \cite{},\footnote{Elliptical distributions have been used in economics since at least \cite{OwenRabinovitch1983} and \cite{Chamberlain1983}.}  and its Gaussian predecessors
there are no fully separating equilibria because preferences violate single-crossing.
\end{comment}
%I analyze scoring in a multi-feature linear-quadratic-elliptical environment in which preferences violate single-crossing. 

Few papers have studied scoring as a form of intermediation. The closest paper is \cite{BonattiCisternasFC}. In their model, a sequence of monopolists price-discriminate using a score that aggregates noisy measurements of past purchase quantities. Scores that overweight the distant past can improve information transmission. The driver of information loss in their model is exogenous noise, not a failure of single-crossing.\footnote{\cite{Segura-Rodriguez2021wp} studies the profit-maximizing sale of scores that aggregate multi-dimensional consumer data. In his model, the consumer data is exogenous. He focuses on screening the data buyer, who has private information.} In single-dimensional signaling games, \cite{Rick2013WP} and \cite{Whitmeyer2019WP} study the effect of garbling the signal over a noisy channel.\footnote{In a moral hazard setting, \cite{Cremer1995} shows  that a principal may prefer a less accurate monitoring technology so that she finds it sequentially optimal to refuse to renegotiate.} 

My screening setting differs from the classical screening literature because the privately informed agent chooses costly actions, rather than cheap talk messages. The contemporaneous paper of \cite{FrankelKartik2019WP} compares screening and signaling in the single-feature LQE setting from \cite{FrankelKartik2019}. They show that the receiver benefits from committing to under-react to the sender's feature in order to discourage sender distortion.\footnote{\cite{DworczakDuffie2018WP} study a specific screening problem: the optimal transaction-size weighting of financial benchmarks in the presence of potential manipulation by traders.}  I will compare my results with theirs throughout the paper. In other models, the receiver faces a binary decision. It may be optimal to commit to an ex-post suboptimal threshold in order to change the sender's incentives for evidence manipulation \citep{Oyarzun2023}, signal-jamming \citep{CunninghamMorenodeBarreda2019WP}, or feature modification in a classification algorithm \citep{Dalvi_etal2004, Hardt_etal2016,Hu_etal2019}.  Finally,  in a field experiment, \cite{Bjorkegren2021wp} estimate the parameters of a screening model similar to mine. They find that the associated optimal screening rule outperforms an alternative rule that is trained without considering strategic manipulation. 
%it changes the incentives to manipulate the evidence or data on the basis of which these decisions are made \, . In computer science, there is a growing literature on such problems of strategic classification  BrucknerScheffer2009,.\footnote{\ \citeconsider strategic classification with heterogeneous distortion ability.  They study how this heterogeneity affects the distribution of welfare across different sender types, while I focus on the informational loss for the receiver.} 

The intermediary in my model controls information about the sender's endogenous features. In the literature on information design, by contrast, the state is generally exogenous. There are a few exceptions. In \citet{BoleslavskyKim2018WP}, \cite{RodinaFarragut2016WP}, \cite{Mekerishvili2018WP}, and \cite{Zapechelnyuk2019WP}, the designer simultaneously persuades the receiver and  motivates the sender to exert effort.\footnote{\cite{Perez-RichetSkreta2018WP} study persuasion with a different model of distortion: the sender can falsify the input to the designer's chosen test.} \cite{Rodina2016WP} and \cite{HornerLambertFC} solve for effort-maximizing feedback in \citeapos{Holmstrom1999} career concerns model.  In my model, the objective is  not to encourage productive effort, but rather to discourage unproductive distortion.

\section{A model of predictive scoring} \label{sec:model}

\subsection{Setting}
%SUMMARY 
There are three players. The agent being scored is called the sender (she). An intermediary (it) commits to a rule that maps the sender’s feature vector into a score. A receiver (he) observes this score and makes a decision. 

% RECEIVER
The receiver wants to predict the sender's \emph{quality} $\th \in \R$.  The receiver makes a decision $y \in \R$, and his utility $u_R$ is given by
\[
	u_R = - (y - \th)^2.
\]
Thus, the receiver matches his decision with his posterior expectation of $\th$.

% SENDER
The sender has $k$ manipulable \emph{features}, labeled $j = 1, \ldots, k$, where $k \geq 1$. For each feature $j$, the sender privately knows her \emph{intrinsic level} $\h_j \in \R$ and her \emph{distortion ability} $\d_j \in \R_+$. (As a mnemonic, $\h$ is for \emph{in}trinsic and $\d$ is for \emph{d}istortion.) The intrinsic levels $\h_j$ are correlated with the quality $\th$---the joint distribution is specified in \cref{sec:LCE}. The sender does not observe her quality $\th$.\footnote{The equilibria that I construct would remain equilibria if the sender observed her quality $\th$.}

For each feature $j$, the sender chooses distortion $d_j \in \R$. Feature $j$ takes the value 
\[
	x_j = \h_j + d_j.
\]
The sender's utility $u_S$ is given by
\[
	u_S 
	=
	y - (1/2) \sum_{j = 1}^{k} d_j^2/ \d_j.
\]
The sender wants the decision $y$ to be high, and she experiences a quadratic cost from distorting each feature. The higher the sender's distortion ability $\d_j$, the lower is her marginal cost of distorting feature $j$. If $\d_j = 0$, the sender cannot distort feature $j$, so $x_j  = \h_j$.%
\footnote{For $\d_j = 0$, the sender's utility is defined by the limit as $\d_j$ converges downward to $0$.} 
%\footnote{The term \textit{feature} comes from statistical learning theory \citep[e.g.,][p.~9]{HastieTibshiraniFriedman2009} but they serve as signals in the sense of \cite{Spence1973}.} 
%Equivalently, this utility function can be expressed in terms of $x_i$ rather than $d_i$. Modeling the distortion choice as primitive  makes it easier to incorporate noise in to the mapping from distortion to features.

% Distributions
%The random $(1 + 2k)$-vector $(\th, \h, \d)$ has an elliptical distribution with finite second moments. Elliptical distributions are more general than the Gaussian. They are flexible enough to accommodate the sign restriction on $\d$, yet retain the convenient property that conditional expectations are linear. This property is stated formally in \cref{sec:LCE}. I make further covariance assumptions in \cref{sec:LCE} and in \cref{sec:existence_uniqueness}.

% INTERMEDIARY
The intermediary, unlike the sender and receiver, has commitment power. Initially, the intermediary commits to a score set $S$ and a scoring rule
\[
	f \colon \R^k \to \D(S),
\]
which assigns to each realized feature vector $x$ a random score in $S$.\footnote{The intermediary in my model directly observes payoff-relevant choices by the sender. By contrast, a classical mediator elicits cheap talk reports \citep{Aumann1974, Myerson1982,Myerson1986, Forges1986}.}
%Here $\D (S)$ is the set of probability measures on $S$, but I use $f(x)$ to denote the random score itself.%
%\footnote{All measurability issues are handled in \cref{sec:measurability}.}
The score set $S$ is not restricted, but I will show that there is no loss in taking $S$ equal to $\R$, the receiver's decision space.  The intermediary has the same utility function as the receiver, so it wants the receiver's decision to be as accurate as possible (in the sense of minimizing mean squared error). An interpretation is that the intermediary is a monopolist who sells a scoring service to the receiver and can extract a fixed share of the receiver's surplus. This is a reasonable approximation of Fair Isaac Corporation (FICO), whose score is required for government-backed mortgages through Fannie Mae and Freddie Mac.\footnote{``Fico’s Dominance in US Credit Scoring under Challenge'' (\href{https://www.ft.com/content/046dd89d-7047-4ed4-af31-0e126f59eb00}{\textit{Financial Times}, 2021}). FICO has a 90\% market share, and it is currently facing antitrust litigation alleging collusion with the three major credit bureaus (Equifax, Experian, TransUnion). Each bureau applies FICO's scoring algorithm to the consumer data that it collects. See \href{https://fingfx.thomsonreuters.com/gfx/legaldocs/zgporjzmzvd/N.D.Ill._1_20-cv-02114_173_0\%20(1).pdf}{Document 173, Case 1:20-CV-02114}. } The receiver's decision can represent, in reduced form, the outcome of a competitive banking market.  
%Even if each individual bank is capable of committing to a decision rule, scoring is a reasonable model if competition between firms makes it unsustainable to take ex-post suboptimal decisions. 

%so the intermediary seeks to minimize the mean squared error of the receiver's decision. 
%between the receiver's decision and dof the receiver's decision Therefore, the intermediary designs the score to make the receiver's decision most accurate (in the sense of minimizing mean squared error).  but this microfoundation is not in the model. 
% allows me to focus on infomration trnamission

% TIMING

% FIGURE:  Flow of information 
\begin{figure}
	\begin{center}
		\begin{tikzpicture}
		
		\tikzstyle{block} = [rectangle,anchor = north, draw, text width=8 em, text centered, rounded corners, minimum height=1.5em]
		
		% nodes 
		\node (N) {};
		\node (S) [block, below = 1cm of N] {\textbf{Sender} \\ distortion $d \in \R^k$};
		\node (R) [block, base right = 4 cm of S] {\textbf{Receiver} \\ decision $y \in \R$};
		\node (I) [block, below right = 3 cm of S.center] {\textbf{Intermediary} \\ commits to\\ $f \colon \R^k \to \D (S)$};
		
		% paths
		\path (N) edge [thick,-{Latex[length=2mm]}] node [anchor = east] {$(\h,\d)$} (S.north);
		\path (S.south) edge [bend right, thick,-{Latex[length=2mm]}] node [below left] {$x = \h + d$} (I.west);
		\path (I.east) edge [bend right, thick,-{Latex[length=2mm]}] node [below right] {$f(x)$} (R.south);
		
		\end{tikzpicture}
	\end{center}
	\caption{Flow of information}
	\label{fig:flow_of_information}
\end{figure}

The intermediary's scoring rule induces a game between the sender and the receiver. \cref{fig:flow_of_information} illustrates the flow of information in this game. The sender observes her private type $(\h, \d)$ and then chooses how much to distort each feature. Her distorted feature vector $x$ is observed by the intermediary. The intermediary assigns the score $f(x)$. The receiver observes the score $f(x)$, but not the feature vector $x$. The receiver updates his beliefs about the sender's quality $\th$ and makes a decision $y$. Finally, payoffs are realized.
%\footnote{The quality $\th$ is not observed by any player in the game, so I could drop $\th$ and replace each player's ex post utility with its expectation conditional on $(\h, \d)$. I find it clearer to explicitly specify the latent quality of interest.}

\subsection{Elliptical distribution} \label{sec:LCE}

The $(1 + 2k)$-vector $(\th, \h, \d)$ follows an elliptical distribution satisfying $\d \geq 0$. 
The second moments are finite. Denote the mean and variance of $(\th, \h, \d)$ by 
\[
\mu = 
\begin{bmatrix} 
\mu_\th \\ 
\mu_\h \\
\mu_\d
\end{bmatrix}
\quad
\text{and}
\quad
\S =
\begin{bmatrix}
\s_\th^2 & \S_{\th \h} & \S_{\th \d} \\
\S_{\h \th} & \S_{\h \h} & \S_{\h \d } \\
\S_{\d \th} & \S_{\d \h} & \S_{\d \d}
\end{bmatrix}.
\] 
The family of elliptical distributions generalizes the multivariate Gaussian. For multivariate Gaussian distributions, (i) the isodensity curves are concentric ellipsoids centered at the mean, and (ii) along any ray emanating from the mean, the density decays exponentially in the squared distance. Elliptical distributions retain property~(i) but relax property~(ii) to allow for any normalized radial density function.%
\footnote{In fact, an elliptical distribution need not have a density. As defined in \cite{CambanisHuangSimons1981}, a random vector  $X$ is elliptically distributed if  there exists a vector $\mu$,  a positive semidefinite matrix $\S$, and a function $\phi$ such that $\E[ e^{it^T(X - \mu)}] =  \phi(t^T \S t)$ for all real vectors $t$.}  In particular, the elliptical family includes distributions satisfying the sign restriction $\d \geq 0$ (which jointly restricts $\mu$, $\S$, and the radial density function). One such example is a uniform distribution on the interior of a suitable ellipsoid. In general, any elliptical distribution can be truncated so that it is supported inside a chosen isodensity ellipsoid. The truncated distribution remains elliptical.

Elliptical distributions are the most general class with the linear conditional expectations property.\footnote{See the monograph of \cite{FangKotzNg1989}.} Denote the regression coefficient vector from regressing a random variable $Y$ on a random vector $X$ by $\reg ( Y | X) = \var^{-1} (X) \Cov(X,Y)$, provided that $X$ and $Y$ have finite second moments and $\var(X)$ has full rank. The next result follows from \citet[Corollary 5,  p.~376]{CambanisHuangSimons1981}.

\begin{lem} [Linear conditional expectations] \label{res:LCE}
	Let $X = A \h + B \d$ for some matrices $A$ and $B$. If $\var(X)$ has full rank, then
	\[
	\E [ \th | X] = \E[\th] + \reg^T(\th | X) (X - \E [X]).
	\]
\end{lem}
In general, regression yields the best \emph{linear} prediction. Here, the regression on the right side is the best \textit{unrestricted}  prediction of $\th$ from $X$ because $(\th,\h, \d)$ is elliptically distributed.

Lastly, I impose standing assumptions on the covariance. 

\begin{as*}[Covariance] \hfill
	\vspace{-\baselineskip}
	\begin{enumerate}[label = A\arabic*., ref = A\arabic*]
		\item \label{it:correlated} $\S_{\h \th} \neq 0$.
		\item \label{it:Schur} $\S_{\h \h} - \S_{\h \d} \S_{\d\d}^{\dagger} \S_{\d \h}$ has full rank.\footnote{Here, $\S_{\d\d}^{\dagger}$ denotes the Moore--Penrose inverse of $\S_{\d\d}$.}
		\item \label{it:covariance}  $\S_{\d \d}$ is componentwise nonnegative. 
		\item \label{it:uncorrelated} $\S_{\d \th} = 0$ and $\S_{\d \h} = 0$.
	\end{enumerate}
\end{as*}

The first two assumptions are technical. Assumption~\ref{it:correlated} ensures that at least one of the sender's intrinsic levels provides relevant information about her quality $\th$. In Assumption~\ref{it:Schur}, the matrix expression is proportional to the conditional variance of $\h$ given $\d$.%
\footnote{For general elliptical distributions, the conditional variance $\var(\h | \d)$ is random, but every realization is a scalar multiple of $\S_{\h \h} - \S_{\h \d} \S_{\d\d}^{\dagger} \S_{\d \h}$  \citep[Corollary 5,  p.~376]{CambanisHuangSimons1981}.} I require this matrix to have full rank so that the distorted feature vector is guaranteed to be nondegenerate. 

The next two assumptions are more substantive. Assumption~\ref{it:covariance} says that the sender's distortion abilities on different features are nonnegatively correlated. This assumption is natural if distortion ability on each feature is a combination of general distortion ability and (uncorrelated) feature-specific distortion ability.  Assumption~\ref{it:uncorrelated} says that the sender's distortion abilities are not correlated with her quality or with her intrinsic levels. This stylized assumption captures the motivation that the sender's intrinsic levels are what convey information about her quality. Together, Assumptions~\ref{it:covariance} and \ref{it:uncorrelated} allow me to obtain an interpretable comparison between the signaling, scoring, and screening solutions, despite the complexity created by the multiple features. 

\section{Signaling without the intermediary} \label{sec:signaling}

For this section, drop the intermediary and suppose that the receiver directly observes the sender's distorted feature vector. The sender and receiver play a signaling game, with features as signals. This signaling setting gives a lower bound on the intermediary's payoff from optimal scoring since full disclosure is a feasible scoring rule.

\subsection{Linear equilibrium characterization} \label{sec:linear_equil_char}

In the signaling game, (pure) strategies are defined as follows. Let $T$ denote the support of $(\h, \d)$. A distortion strategy for the sender is a map $d \colon T \to \R^k$, which assigns a distortion vector to each sender type. A decision strategy for the receiver is a map $y \colon \R^k \to \R$, which assigns a decision to each feature vector.

I focus on \emph{linear equilibria}, i.e., Bayes--Nash equilibria in which each player's strategy is linear (technically affine).%
\footnote{The sign restriction $\d \geq 0$, combined with the elliptical symmetry of $(\h, \d)$, forces $(\h, \d)$ to have bounded support. Hence, some feature vectors are off-path. In \cref{sec:PBE}, I discuss how the linear equilibrium outcome can be achieved as a perfect Bayesian equilibrium.}  Linear equilibria demand less sophistication from the players---running regressions is sufficient. Moreover, decisions made according to linear rules can be easily explained by specifying the weight placed on each feature. Finally,  linear equilibria are tractable due to the linear conditional expectations property of elliptical distributions.\footnote{By contrast, nonlinear equilibria require difficult computations of expected quality, conditional on the sender's type lying in particular subsets of $\R^{2k}$.} The linearity assumption is restrictive, however. In \cref{sec:nonlinear_equilibria}, I construct a family of (unrealistic) nonlinear Bayes--Nash equilibria in the single-feature setting.  If distortion ability is sufficiently heterogeneous, then there exists a nonlinear equilibrium that the receiver strictly prefers to the linear equilibrium.

%With multi-dimensional types, signaling equilibria are generally difficult to construct, unless we focus on equilibrium in linear strategies or on discrete (often binary) action spaces. Discrete action spaces suffer from a ceiling effect: When types pool on a high action, there is a mechanical loss of information. To avoid this effect, I focus on a continuous model with linear equilibria. Requiring off-path beliefs to lie in the support means that some type would not distort her feature. This low probability event makes belief-updating intractable, so I instead user linearity it discipline off-path beliefs The linear conditional expectation property of the elliptical distribution makes these equilibria tractable. 

To analyze linear equilibria, suppose that the receiver uses the linear strategy 
\begin{equation*} 
	y(x) = b_0 + b^T x,
\end{equation*}
for some intercept $b_0$ and coefficient vector $b = (b_1, \ldots, b_k)$.  Given this strategy, the sender's utility from distortion vector $d$ is 
\[
	b_0 + b^T (\h + d) - (1/2) \sum_{j =1}^{k} d_j^2 /\d_j. 
\]
This expression is strictly concave in $d$ and additively separable across features. The marginal benefit of distorting feature $j$ equals  $b_j$; the marginal cost equals $d_j/\d_j$. Equating these expressions yields the sender's best response: $d_j (\h, \d ) = b_j \d_j$ for each $j$.  That is, $d(\h, \d)$ is the componentwise product $ b \circ \d$. 

If the sender plays this best response $d(\h, \d) = b \circ \d$, then her feature vector equals $\h + b \circ \d$. The equilibrium condition is
\begin{equation} \label{eq:signaling_CE}
b_0 + b^T ( \h + b \circ \d) = \expec [ \th |\h + b \circ \d ].
\end{equation}
This is an equality between random variables. The left side is the  decision specified by the receiver's strategy at the sender's feature vector. The right side is the receiver's posterior expectation of $\th$ upon seeing the sender's feature vector. The intercept $b_0$ does not affect the sender's incentives, and it is pinned down by $b$ (since  the expected decision must equal $\mu_\th$). Hereafter I mention only the coefficient vector $b$.  By the linear conditional expectations property (\cref{res:LCE}), the linear equilibrium condition reduces to the regression equation
\begin{equation} \label{eq:signaling_reg}
	b = \reg(\th | \h + b \circ \d).
\end{equation}
The variance of $\h + b \circ \d$ is quadratic in $b$, so  \eqref{eq:signaling_reg} is cubic.%
\footnote{This contrasts with other linear-quadratic settings. In classical linear-quadratic coordination games \citep{MorrisShin2002} and their generalization in  \cite{LambertMartiniOstrovsky2018WP}, the equilibrium condition is linear because signals are exogenous. In \cite{Kyle1985}, the signal is endogenous, but the sensitivity of the sender's (insider's) best response is \emph{inversely} proportional to the sensitivity of the receiver's (market maker's) strategy. As a result, the equilibrium condition is quadratic.}
Nevertheless, the covariance assumptions guarantee that \eqref{eq:signaling_reg} has a unique solution.

\begin{thm}[Existence and uniqueness] \label{res:existence_uniqueness}
	The signaling game has exactly one linear equilibrium.
\end{thm}

%Denote the unique equilibrium vector by $b^{\signal}$. (Below, the solutions in the scoring and screening settings will be denoted $b^{\score}$ and $b^{\screen}$.) 

The only covariance assumption required for  existence is \ref{it:Schur}, which gives a uniform lower bound on $\var(\h + b \circ \d)$. Thus, the regression vector $\reg (\th | \h + b \circ \d)$ is bounded as a function of $b$, so existence follows from Brouwer's fixed point theorem.

Uniqueness uses the covariance assumptions \ref{it:covariance} and \ref{it:uncorrelated}. The intuition is clearest with a single feature ($k = 1$).\footnote{In the single-feature setting, uniqueness is shown in \citet[Proposition 1, p.~236]{FischerVerrecchia2000}.  \citet[Proposition 4, p.~1761]{FrankelKartik2019} extend uniqueness to the case in which $\s_{\h \d} \geq 0$.  If $\s_{\h \d} < 0$, they observe that there can be multiple increasing linear equilibria. \citet[Proposition 1b, p.~459]{Gesche2021} analyzes uniqueness in a variant of the single-feature model in which the receiver is naive with some fixed probability.} In that case, the equilibrium condition \eqref{eq:signaling_reg} becomes
\[
		b  = \frac{\s_{\th \h}}{\s_{\h}^2  + b^2 \s_{\d}^2},
\]
where I use $\s$ in place of $\S$ for scalar covariances. Consider an equilibrium with coefficient $b^\ast$. Suppose $\s_{\th \h} > 0$. Thus, $b^\ast > 0$. If the receiver increases the coefficient from $b^\ast$ to $b$, then the sender's best response is to distort her feature more. This makes her feature less informative, so the receiver's best response is \emph{less} sensitive:
\[
	\reg ( \th | \h +b \circ \d) < \reg ( \th | \h +b^\ast \circ \d)  = b^\ast < b.
\]
Symmetrically, if the receiver reduces the coefficient from $b^\ast$ to some $b$ in $(0, b^\ast)$, then the sender's best response is to distort less, making the receiver's best response \emph{more} sensitive. If there are multiple uncorrelated features, then this argument can be applied componentwise.

%Therefore, the equilibrium condition \eqref{eq:signaling_reg} is a cubic system of multivariate polynomial equations.

In  the general case with multiple correlated features, the regression coefficient on one feature depends on the sender's distortion on all features.  Using Assumptions \ref{it:covariance} and \ref{it:uncorrelated},  I show that a coefficient vector induces a linear equilibrium if and only if it minimizes the function
\[
	\Phi(b) =  \var ( b^T  \h   - \th) + (1/2) \var ( b^T (b \circ \d)).
\]
This function is strictly convex, so it has at most one minimizer. The function $\Phi$ is similar to a potential function \citep{Rosenthal1973,MondererShapley1996}.\footnote{Recall that a best-response potential game is best-response equivalent to an \emph{identical interest} game; see \cite{Voorneveld2000}. Here, if the sender and receiver are restricted to linear strategies, then the signaling game is best-response equivalent to a \emph{zero-sum} game. In this zero-sum game, each player's utility function is concave, so it can be shown that continuous best-response dynamics converge to the unique linear equilibrium; see \cite{BarronGoebelJensen2010}, which generalizes \cite{HofbauerSorin2006}.}  The factor of $1/2$ in the second term of $\Phi$ isolates the marginal effect of 
the \emph{receiver's} strategy on the squared decision error; the other half is due to the \emph{sender's} best response.

\subsection{Information loss from distortion}

Let $\b$ denote the regression vector $\reg ( \th | \h)$. If the receiver could observe the sender's type $(\h, \d)$, he would form the posterior expectation
\[
	\E [ \th | \h, \d ] = \b_0 + \b^T \h,
\]
where $\b_0 = \mu_\th - \b^T \mu_\h$. An equilibrium is \emph{fully informative} if the receiver's decision coincides with $\E [ \th | \h, \d]$.\footnote{Redefining $\th$ to equal $\E [ \th | \h, \d]$ would not change my analysis. It would only translate the utilities of the intermediary and the receiver.}

%During the game, no player observes the residual uncertainty $\th - \E [ \th |\h, \d]$, so quality $\th$ could be redefined to $\b_0 + \b^T \h$ without changing behavior.

\begin{prop}[Fully informative equilibrium] \label{res:fully_informative_linear}
	In the signaling game, the unique linear equilibrium is fully informative if and only if  $(\b \circ \b)^T \d$ is constant, i.e., $(\b \circ \b) ^T\S_{\d\d} (\b \circ \b) = 0$. 
\end{prop}

If the distortion ability vector $\d$ is homogeneous in the population, then the equilibrium is fully informative.  Without uncertainty about the sender's distortion ability, the sender's distortion best response can be perfectly anticipated and subtracted to recover the intrinsic levels. Even if  the distortion best response  $\b \circ \d$ to the coefficient vector $\b$ cannot be perfectly anticipated, as long as its contribution $\b^T ( \b \circ \d)$ to the decision can be anticipated, there is no information loss.

If the signaling equilibrium is fully informative, then in the scoring setting the intermediary can do no better than fully disclosing the sender's feature vector. Therefore, I am primarily interested in the case in which $\S_{\d\d}$ has full rank. In this case, \cref{res:fully_informative_linear} guarantees that the unique linear equilibrium in the signaling game is not fully informative. In fact, there is no fully informative Bayes--Nash equilibrium, linear or otherwise; see \cref{sec:beyond_linear}.

Next, I study how the  population heterogeneity of the distortion ability vector $\d$ affects the informativeness of the equilibrium. Scaling up the distortion variance $\S_{\d\d}$  is equivalent to increasing the weight that the sender places on the decision. Formally, following \cite{FrankelKartik2019}, suppose that the sender's utility equals
\[
	\a y - (1/2) \sum_{j = 1}^{k} d_j^2/\d_j,
\]
where $\a > 0$. Call $\a$ the sender's \emph{stake} in the decision. Dividing the sender's utility by $\a$ does not change her preferences, so the model with stake $\a$ is equivalent to the baseline model with distortion ability $\a \d$ in place of $\d$, or equivalently, distortion variance $\a^2 \S_{\d\d}$ in place of $\S_{\d\d}$. 

\begin{prop}[Sender's stake and information loss] \label{res:info_loss}
If $\S_{\d\d}$ has full rank, then the receiver's signaling utility is strictly decreasing in the sender's stake.
\end{prop}

That is, decision accuracy is decreasing in the sender's stake. Fixing a strategy for the receiver, it is clear that with a higher stake the sender's best response adds more noise to the feature vector, reducing decision accuracy. But with a higher stake, the receiver's equilibrium strategy also changes.  I show that the change in the receiver's strategy is not large enough to offset the direct effect of the increased stake. 

 \citet[Proposition 4, p.~1761]{FrankelKartik2019} establish the same conclusion as \cref{res:info_loss} in their single-feature  signaling setting.\footnote{Their result allows for nonnnegative correlation between the agent's intrinsic level (which equals quality) and distortion ability. \citet[Corollary 1, p.~238--239]{FischerVerrecchia2000} give an equivalent result, assuming zero correlation. To keep my multi-feature model tractable, I rule out correlation between the intrinsic levels and the distortion abilities; see Assumption~\ref{it:uncorrelated}.} With a single feature, scaling up $\S_{\d\d}$ is equivalent to increasing $\S_{\d\d}$ in the positive semidefinite order. With multiple features, this equivalence no long holds. In fact,  increasing $\S_{\d \d}$ in the positive semidefinite order does not necessarily reduce the receiver's equilibrium utility; see \cref{sec:counterexamples} for an example. 
 %which is equivalent to scaling up $\S_{\d \d}$ if $k = 1$

 %(with nonnegative correlation between $\th$ and $\h$)
 
 %\cref{res:info_loss} only covers the case $\a_1 = \cdots = \a_k$. If the $\a_i$ are very unequal, the change in the receiver's equilibrium strategy can more than offset the direct effect of more heterogeneous distortion. 

%In their setting, increasing stakes is equivalent to increasing the variance of distortion ability. In my multidimensional setting, there are two natural extensions: scaling up $\S_{\d\d}$ or increasing $\S_{\d\d}$ in the positive semidefinite order. \cref{res:info_loss} shows that the weaker scaling extensions holds; \ shows, however, that the stronger order extension fails. 

%With multiple features, the variance matrix can be increased in different directions that do not correspond to increasing the stakes. My result pinpoints the increasing stakes as the source of  reduced equilibrium information transmission. Fixing the receiver's strategy, The informativeness of the sender's features are indeed decreasing in $\S_{\d\d}$, but the equilibrium effect on the receiver's strategy can outweigh this direct effect, as shown in the following counterexample. 

\section{Scoring} \label{sec:scoring}

Now I return to the scoring model with the intermediary. I first characterize the linear decision rules that the intermediary can induce. Then I maximize the intermediary's objective over this set, and I analyze the optimal weight placed on each feature.  

%Using the revelation principle, To simplify the intermediary's problem, I first establish a revelation principle: There is no loss in restricting scores to direct decision recommendations that the receiver obeys. This is not a standard revelation principle because the intermediary observes costly actions, not cheap-talk reports. 

\subsection{Revelation principle and obedience}

Each scoring rule $f \colon \R^k \to \D(S)$ induces a game between the sender and the receiver. In this game, the sender chooses a (mixed) strategy $d \colon T \to \D(\R^k)$ and the receiver chooses a (mixed) strategy $y \colon S \to \D(\R)$. The intermediary's score set $S$ is unrestricted, but the logic of the revelation principle from \cite{Myerson1986} shows that there is no loss in restricting to direct, obedient scoring rules. That is, scores are decision recommendations that the receiver is willing to obey.

%\footnote{Here is the formal argument. Let $\id$ denote the identity function. If $(d,y)$ is a Bayes--Nash equilibrium of the game induced by a scoring rule $f$, then $(d, \id)$ is a Bayes--Nash equilibrium of the game induced by the direct scoring rule $y \circ f \colon \R^k \to \D(\R)$.}

As in the signaling benchmark, I restrict attention to equilibria in which the receiver's decision is a linear function of the sender's feature vector. Formally, I restrict attention to scoring rule--equilibrium pairs $(f; d,y)$ for which $y \circ f$ is a deterministic linear function from $\R^k$ to $\R$.\footnote{If random scoring is allowed, but the \emph{expected} score is required to be a  linear function of the features, then randomness is suboptimal. See \cref{sec:extensions}.}  This restriction keeps the problem tractable, and it ensures that the comparison between scoring and signaling isolates the effect of intermediation (rather than nonlinearity). 

Once I impose this substantive linearity restriction, the revelation principle says that there is no \emph{further} loss in restricting attention to obedient scoring rules $f \colon \R^k \to \R$ of the form 
\[
	f(x) = b_0 + b^T x,
\]
for some intercept $b_0$ and coefficient vector $b = (b_1, \ldots, b_k)$. As in the signaling game, the sender's best response is  $d(\h, \d) = b \circ \d$. The receiver's obedience condition is 
\begin{equation} \label{eq:scoring_CE}
	b_0 + b^T ( \h + b \circ \d) = \E [ \th | b_0 + b^T ( \h + b \circ \d)].
\end{equation}
The right side is the receiver's posterior expectation of $\th$ upon seeing the score. Condition \eqref{eq:scoring_CE} is weaker than the signaling equilibrium condition \eqref{eq:signaling_CE} because the receiver's posterior expectation in  \eqref{eq:scoring_CE} is based upon less information. Suppose that there are multiple features ($k > 1$). In feature space $\R^k$, the level sets of the scoring rule are parallel hyperplanes normal to the coefficient vector $b$. The receiver forms his posterior expectation knowing only the hyperplane on which the sender's feature vector lies.

By the linear conditional expectations property (\cref{res:LCE}), condition \eqref{eq:scoring_CE} holds if and only if either $b = 0$ or else $b \neq 0$ and 
\begin{equation} \label{eq:scoring_reg}
	1 = \reg (\th | b^T ( \h + b \circ \d)),
\end{equation}
where $b_0$ is not mentioned since it is pinned down by $b$. Thus, obedience requires that the regression coefficient on the score is $1$. By contrast, the signaling equilibrium condition matches the regression coefficients on all $k$ features.\footnote{With a single feature ($k = 1$), conditions  \eqref{eq:scoring_reg} and \eqref{eq:signaling_reg} are equivalent. In this case, the only difference between scoring and signaling is that scoring allows $b = 0$, i.e., providing no information, but this will never be optimal.}

\subsection{Optimal scoring} \label{sec:optimal_scoring}

%So far, I have characterized the set of linear obedient decision rules that the intermediary can induce. The analysis so far holds for any objective of the intermediary. Now I maximize the receiver's utility over this set, and I will identify when the intermediary can strictly improve the receiver's payoff. 

%\cref{sec:efficient_scoring} considers a more general social welfare objective. 

I maximize the intermediary's objective over the space of obedient linear scoring rules. Obedience implies that the mean decision error is zero, so the mean squared error reduces to the variance. After reformulating the obedience condition \eqref{eq:scoring_CE} in terms of covariances, the intermediary's problem becomes
\begin{equation} \label{eq:scoring_problem}
\begin{aligned}
	& \text{minimize} 
	&& \var \bigl( b^T (\h +  b \circ \d) - \th \bigr) \\
	& \text{subject to } 
	&&    \var \bigl( b^T(\h + b \circ \d) \bigr)
			=
			\cov \bigl( b^T (\h + b \circ \d), \th \bigr).
\end{aligned}
\end{equation}

%The objective is continuous and strictly convex, and it has compact sublevel sets.  The constraint set is closed, but generally not convex. Nevertheless, by analyzing the Lagrangian, I establish uniqueness. 

\begin{comment}
\footnote{The bias--variance decomposition gives
\[
\E^2 \bigl[ b_0 + b^T (\h + b \circ \d) - \th \bigr] + \var \bigr( b_0 + b^T (\h + b \circ \d) - \th \bigr).
\]
For obedient decision rules, the first term vanishes. In the second term, $b_0$ does not affect the variance so it can be dropped.
}
\end{comment}

% To illustrate the return from commitment, suppose that the receiver commits to perturb his strategy away from his equilibrium strategy. Suppose also that the sender plays a best response to this perturbation. Starting at an equilibrium, neither player's perturbation has a first-order effect on his or her own payoff. But the sender's perturbation does have a first-order effect on the receiver's payoff.

It can be shown that the scoring problem \eqref{eq:scoring_problem} has a unique solution,\footnote{The argument is given in the proof of \cref{res:scoring_equals_signaling} (\cref{sec:proof_scoring_equals_signaling}).} which I denote by $b^{\score}$. Let $b^{\signal}$ denote the coefficient vector in the unique linear signaling equilibrium. The vector $b^{\signal}$ is a feasible point in \eqref{eq:scoring_problem}, but it is generally suboptimal.

%Perturbing the coefficient vector $Deb$ away from $b^{\signal}$ has two effects on the mean-square error. First, it introduces ex-post prediction error---the receiver's decision is not the optimal prediction of quality from the sender's distorted features. Second, it induces the sender to choose a different level of distortion, which changes the distribution of the sender's features. At  $b^{\signal}$, the ex-post prediction error is minimized, so near $b^{\signal}$ the loss from prediction error is only second order. As long as the equilibrium is fully informative, the gain from inducing more informative signals is first order. Therefore, there exists a strictly improving pertuthere is an \emph{obedient} strictly improving perturbation, unless the gradient of the receiver's utility at $b^{\signal}$ is orthogonal to the obedience constraint set. This orthogonality condition is formalized in the next theorem. 
%\footnote{\cite{FrankelKartik2019WP} use this first-order reasoning, but in their screening problem there is no obedience constraint.}
%But in my setting, it's actually even easier to show that screeening can strictly improve upon scoring. 
% 

\begin{thm}[Scoring versus signaling] \label{res:scoring_equals_signaling}
The following are equivalent:
\begin{enumerate}[label = (\roman*), ref= \roman*]
	\item \label{it:coincide} $b^{\signal} = b^{\score}$;
	\item \label{it:scale} $b^{\signal}$ is a scalar multiple of $\b$;
	\item \label{it:parameter} $\diag(\b) \S_{\d \d} (\b \circ \b)$ is a scalar multiple of  $\S_{\h \th}$.
\end{enumerate}
\end{thm}

Recall that  $b^{\signal} = \reg ( \th | \h + b^{\signal} \circ \d)$ and $\b = \reg ( \th | \h)$. Thus, condition~\ref{it:scale} says that 
\[
	\reg(\th | \h + b^{\signal} \circ \d) = \l \reg ( \th | \h),
\]
for some scalar $\l$ (which will necessarily be between $0$ and $1$). This means that in the signaling equilibrium, the sender's distortion dampens the informativeness of every feature by the same factor. Thus, condition \ref{it:scale} is a symmetry condition. Expressed in terms of primitives, condition~\ref{it:scale} reduces to condition~\ref{it:parameter}.

With multiple features, the symmetry condition is a knife-edge case. In view of condition~\ref{it:parameter}, if $k>1$ then for a generic set of covariance parameters $( \S_{\h\h}, \S_{\h \th},\S_{\d\d})$, scoring strictly improves upon signaling. That is, the intermediary can adjust the coefficients in the signaling equilibrium  to strictly increase information transmission, while preserving obedience. With a single feature, the symmetry condition holds automatically, so signaling and scoring coincide. Indeed, any nontrivial linear function of a single feature fully discloses the feature.\footnote{This equivalence relies on the restriction to linear rules. Even with a single feature, nonlinear rules could partially garble the sender's feature.}

%Note, however, that the relations $A \succeq B$ and $A \neq B$ do not imply $A \succ B$. 

\begin{prop}[Scoring comparative statics] \label{res:scoring_CS}
The receiver's scoring utility is decreasing in the variance $\S_{\d \d}$.\footnote{\label{ft:decreasing}Let $\succeq$ and $\succ$ denote the positive semidefinite and positive definite matrix orders. A real-valued function $h$ acting on symmetric square matrices is \emph{decreasing} if  (i) $A \succeq B$ implies $h(A) \leq h(B)$, and (ii) $A \succ B$ implies $h(A) < h(B)$.}
\end{prop}

This is the intuitive result that failed for the signaling setting; see \cref{sec:counterexamples}. Consider any obedient scoring rule. If the variance $\S_{\d \d}$ decreases in the positive definite order, then the original scoring coefficient vector can be scaled up until it satisfies the new obedience condition. It can be shown that this procedure strictly increases the receiver's expected utility. 

% In fact, the proof shows that the receiver's payoff is decreasing with respect to the even weaker copositive matrix order. % One example of a matrix in that order is $\d_0$. 

\subsection{Setting with uncorrelated errors} \label{sec:simple_setting}

In order to compare  scoring and signaling feature-by-feature, I consider the following setting of \emph{uncorrelated errors}, which is a special case of the main model. The sender's quality $\th$ is normalized to have unit variance. The intrinsic level of each feature $j$ is given by 
\[
	\h_j = \th + \e_j, 
\]
where the errors $\e_1, \ldots, \e_k$ are uncorrelated.\footnote{The uncorrelated errors assumption means that within any subpopulation of senders who have the same quality $\th$, the intrinsic levels of different features are uncorrelated.} The vector $(\th, \h, \d)$ is determined by the vector $(\th, \e_1, \ldots, \e_k, \d_1, \ldots, \d_k)$, which follows an elliptical distribution with uncorrelated components, satisfying $\d \geq 0$. This setting is parameterized by the variances 
$\s_{\e,j}^2 = \var(\e_j)$ and $\s_{\d,j}^2 = \var(\d_j)$ for $j = 1, \ldots, k$. For each $j$, assume that $\s_{\e,j}^2  > 0$ and $\s_{\d,j}^2  > 0$. For each feature $j$, the parameter $\s_{\e,j}$ controls the imprecision of the intrinsic level, and $\s_{\d,j}$ controls the heterogeneity of distortion ability.

%FIGURE: SCORING v. SCREENING
\begin{figure}
\centering
\begin{tikzpicture}
\begin{groupplot}[	group style = {group size = 2 by 2, horizontal sep = \plotgap}]

	\nextgroupplot[
	axis lines = center,
	scale = 1.1,
	axis equal image,
	xtick = \empty,
	ytick = \empty,
	xmin = -0.2,
	xmax = 0.55,
	ymin = -0.2,
	ymax = 0.55, 
	xlabel = {$b_1$},
	ylabel = {$b_2$},
	x label style={at={(current axis.right of origin)},anchor=west},	
	y label style={at={(current axis.above origin)},anchor=south},	
	]
	
	\addplot [Blue, thick, smooth] table [col sep = comma, x index = 0, y index = 1] {feasible_scoring_symmetric.tex};
	\addplot [Orange, thick, dashed, smooth] table [col sep = comma, x index = 0, y index = 1] {feasible_scoring_asymmetric.tex};
	
	\nextgroupplot[
	xtick = \empty,
	ytick = \empty,
	axis equal image,
	xmin = 0.25,
	xmax = 0.40,
	ymin = 0.25,
	ymax = 0.38, 
	xlabel = {$b_1$},
	ylabel = {$b_2$},
	y label style={rotate = -90},
	scale = 1.1
	]
	
	\addplot [Blue, thick, smooth] table [col sep = comma, x index = 0, y index = 1] {feasible_scoring_symmetric.tex};
	\addplot [Orange, thick, dashed, smooth] table [col sep = comma, x index = 0, y index = 1] {feasible_scoring_asymmetric.tex};
	
	\fill  (1/3,1/3) circle (2pt);
	\addplot [dotted, gray, thick] coordinates {(0.2,0.2) (1/3,1/3)} node [anchor = south] {$\b$};
	\fill [Blue]  (0.322185, 0.322185) circle (2pt) node [anchor = west, xshift = 5pt, yshift = -1pt] {$b^{\signal} = b^{\score}$};
	\fill [Orange]  (0.338292, 0.284701) circle (2pt) node [anchor = east, xshift = -4pt] {$b^{\signal}$};
	\fill [Orange] (0.350345, 0.273207) circle (2pt) node [anchor = east, xshift = -4pt] {$b^{\score}$};
	
\end{groupplot}
\end{tikzpicture}
	\caption{\emph{Left}: Obedient coefficient vectors for two different parameter values. \emph{Right}: Signaling equilibrium and optimal scoring rule.}
	\label{fig:obedient_b}
\end{figure}

\begin{comment}
\begin\begin{wideitemize}
\item less information for the receiver means fewer opportunities to deviate, i.e., set of strategies available to the receiver is smaller
\item effectively constrained to selecting a coefficient vector that is aligned with the vector $b$, just chooses coefficient, but this is not quite true if there is noise
\item run regression on one coefficient instead of $k$
\end\begin{wideitemize}
\end{comment}

As a running example, suppose there are two features, with $\s_{\e,1} = \s_{\e,2} = 1$ and $\s_{\d,1} = 1$. The final parameter $\s_{\d,2}$ will vary. The left panel of \cref{fig:obedient_b} plots the set of obedient coefficient vectors for the symmetric case $\s_{\d,2}  = 1$  (solid blue) and an asymmetric case $\s_{\d,2} = 2$ (dashed orange). 
%Recall that the intercept $b_0$ is pinned down by $b$, and hence it is omitted. 
%In both cases, the obedience curve encloses a convex set. 
Both curves pass through the origin, which corresponds to the completely uninformative scoring rule. As $\s_{\d,2}$ increases from $1$ to $2$, distortion ability becomes more heterogeneous and the obedience curve shrinks inward. If the intermediary uses a scoring rule with a coefficient vector inside the obedience curve, and the sender best responds, then the receiver's best response is a linear function of the score with slope strictly greater than $1$. Conversely, if the intermediary's coefficient vector is outside the curve, and the sender best responds, then the receiver's best response is a linear function of the score with slope strictly less than $1$ (and possibly negative). 

The right panel of 	\cref{fig:obedient_b} magnifies a portion of the left panel. I show the vectors $b^{\signal}$ and $b^{\score}$ (for both sets of parameters) and the regression vector $\b = (1/3, 1/3)$, which does not depend on $(\s_{\d, 1}, \s_{\d,2})$. In the symmetric case (solid blue), $b^{\signal}$ is a scalar multiple of $\b$. In this case, $b^{\signal}$ maximizes the receiver's utility over all obedient decision rules, so $b^{\signal} = b^{\score}$. In the asymmetric case (dashed orange), $b^{\signal}$ puts more weight on the first feature than the second because distortion ability is more heterogeneous on the second feature. Starting from the signaling equilibrium, the intermediary can improve decision accuracy by sliding the coefficient vector along the obedience curve away from the $45$-degree line, shifting even more weight onto the first feature. 

%In general, there exists a local improvement away from the signaling equilibrium, unless the gradient of the receiver's objective is orthogonal to the surface of obedient decision rules. 

For each coefficient vector $b$,  define the weighting ratios
\[
	w_j(b) = \frac{b_j}{\reg_j ( \th | \h + b \circ \d)}, \qquad j = 1, \ldots, n.
\]
Thus, $w_j(b)$ is the ratio between the weight $b_j$ placed on feature $j$ and the weight $\reg_j( \th | \h + b \circ \d)$ that is ex-post optimal, given the induced distortion by the sender.\footnote{I only evaluate the weighting ratio at coefficient vectors $b$ for which  $\reg_j( \th | \h + b \circ \d) > 0$ for all $j$.} A coefficient vector $b$ \emph{underweights} feature $j$ if $w_j (b) < 1$ and \emph{overweights} feature $j$ if $w_j (b) > 1$. In particular, $w_j (b^{\signal}) = 1$ for all features $j$ (and this is an equivalent formulation of the equilibrium condition).  Let $\tau_j$ denote the ratio $\s_{\e,j}^3 / \s_{\d,j}$.

\begin{thm}[Scoring weights] \label{res:underweighting_overweighting}
	In the setting of uncorrelated errors, the following hold.
\begin{enumerate}[label = (\roman*), ref= \roman*]
	\item \label{it:scoring_components} For $\hat{b} = b^{\signal}$ and $\hat{b} = b^{\score}$, the components of $\hat{b}$ are all strictly positive, and $\hat{b}_i > \hat{b}_j$ if $(\s_{\e,i}, \s_{\d, i}) < (\s_{\e,j}, \s_{\d, j})$.\footnote{Here $x < y$ means that $x \leq y$ and $x \neq y$.}
	\item \label{it:scoring_comparison}  $b^{\score} = b^{\signal}$ if and only if  $\tau_1 = \cdots  = \tau_k$. 
	\item \label{it:scoring_weighting} If $b^{\score} \neq b^{\signal}$, then $b^{\score}$ underweights at least one feature and overweights at least one feature. 
	\item \label{it:scoring_ratios} $w_i (b^{\score}) \geq w_j (b^{\score})$ if and only if $\tau_i \geq \tau_j$.
\end{enumerate}
\end{thm}

%	\item $b^{\signal}$ is strictly positive and $b_i^{\signal} > b_j^{\signal}$ if $(\s_{\e,i}, \s_{\d,i}) < (\s_{\e,j}, \s_{\d,j})$.
%\item $b^{\score}$ is strictly positive and $b_i^{\score} > b_j^{\score}$ if $(\s_{\e,i}, \s_{\d,i}) < (\s_{\e,j}, \s_{\d,j})$.

%It would be natural to compare b^{score} and b^{signal}, but I  don't think the relationship between them is clear. Perhaps a referee will ask for a counterexample. 

Consider part~\ref{it:scoring_components}. First, signaling and scoring both put strictly positive weight on every feature. This is intuitive since the intrinsic levels are all positively correlated with the sender's quality.\footnote{With correlated errors, however, the scoring solution could put negative weights on some features in order to offset the contribution of the error terms from other features.} Second, within the same commitment regime (scoring or signaling), the weight on a feature is higher if the intrinsic level is more accurate or the distortion ability is more homogeneous. 
 
 Part \ref{it:scoring_comparison} provides a simple representation of the symmetry condition from \cref{res:scoring_equals_signaling}: the ratio $\tau_j$ is the same for every feature $j$.  If this condition is violated, then $b^{\score}$ is not an equilibrium, i.e., 
 $w_j(b^{\score}) \neq 1$ for some feature $j$.  Obedience then implies that
some feature is underweighted and some feature is overweighted (part \ref{it:scoring_weighting}).
%Of course, it is possible to combine underweighting with overweighting only if there are multiple features. 

%With a single feature, obedience reduces to the signaling equilibrium condition, so $b^{\score} = b^{\signal}$. 
 
Part \ref{it:scoring_ratios} gives an exact ordinal characterization of underweighting and overweighting. The weighting ratios are ordered in the same way as $\tau_j = \s_{\e,j}^3/\s_{\d,j}$.\footnote{The assumption of uncorrelated errors allows me to make feature-by-feature comparisons that depend only on the variance parameters of each feature. Without this assumption, the relationship between $b^{\signal}$ and $b^{\score}$ could depend on all pairwise error correlations.}   Thus, feature $j$ is more overweighted if $\s_{\e,j}$ is larger or $\s_{\d,j}$ is  smaller. To be sure, the \emph{absolute} weight on feature $j$ is decreasing in both $\s_{\e,j}$ and $\s_{\d,j}$ (by part \ref{it:scoring_components}). But the weighting ratio considers the feature weight \emph{relative} to the ex-post optimal weight. 

To build intuition for part \ref{it:scoring_ratios}, recall that the contribution of feature $j$ to the decision is 
\begin{equation} \label{eq:contribution}
	b_j x_j  = b_j  (\th + \e_j + b_j \d_j). 
\end{equation}
Thus, $b_j$ enters directly as the coefficient on $x_j$ and indirectly through the sender's distortion best response. The scoring coefficient and the regression coefficient both take into account that the coefficient on $x_j$ picks up the noise $\e_j + b_j \d_j$ in $x_j$. But only the scoring coefficient internalizes the indirect effect of the coefficient $b_j$  through the sender's distortion best response $b_j \d_j$.  The noise created by this indirect effect drives underweighting.  Feature $j$ is more underweighted if $\s_{\d,j}$ is large relative to $\s_{\e,j}$, i.e., if  $\tau_j = \s_{\e,j}^3/ \s_{\d,j}$ is small. The parameter  $\tau_j$  is more multiplicatively sensitive to $\s_{\e,j}$ than to $\s_{\d,j}$ because of the following dampening effect.  If either $\s_{\e,j}$ or $\s_{\d,j}$ changes, the scoring coefficient $b_j$ moves in the opposite direction (by part \ref{it:scoring_components}), dampening the marginal effect of $b_j$ on the variances of  the terms $b_j \e_j$ and $b_j^2 \d_j$ in \eqref{eq:contribution}. Indeed, $(\mathrm{d}/\mathrm{d} b_j)\var ( b_j \e_j) = 2 b_j \s_{\e,j}^2$ and $(\mathrm{d}/\mathrm{d} b_j) \var ( b_j^2\d_j) = 4 b_j^3 \s_{\d,j}^2$, so the change in $\s_{\d,j}$ is dampened more and the relevant ratio is cubic.

%Why does $\s_{\e,j}$ enter as a cubic? Algebraically, the first-order condition involves the derivative of a quartic expression, which is cubic.   The fundamental intuition is that

%The two terms in parentheses capture the noise introduced into the decision. The ex-post regression weight internalizes the first noise term but only half of the second (the other half is from the sender's distortion best response). The more the second term dominates the variance of this expression, the greater the underweighting.

%Finally, $\tau_j$ is more sensitive to $\s_{\e,j}$ than $\s_{\d,j}$ (through the cubic power) because t
%The dependence on $\s_{\d,j}$ is intuitive---the less heterogeneous is distortion ability on a feature, the smaller is the benefit of underweighting it. The dependence on $\s_{\e, j}$ is more subtle. If the intrinsic level of a feature is a noisier measure of quality, that feature is more \emph{over}weighted. To be sure, as $\s_{\e,j}$ increases, the weight $b_j^{\score}$ decreases (), but the ex post optimal weight decreases even more, resulting in more overweighting.  

\begin{figure}
	\centering

\begin{tikzpicture}
	\begin{axis}[
		clip = false,
		width= 0.85 \textwidth,
		height=0.50\textwidth,
		axis y line=middle, 
		axis x line=bottom,
		clip = false,
		scale = 1.1,
		xtick = {1,4},
		xticklabels = {1,2},
		ytick = {0.333},
		yticklabels = { $\frac{1}{3}$},
		xmin = 1,
		xmax = 4.1,
		ymin = 0.26,
		ymax = 0.36,
		xlabel = {$\s_{\d,2}$},
		x label style={at={(current axis.right of origin)},anchor=west},
		y label style = {at={(current axis.above origin)},anchor=south},
		]

			\addplot [thick] coordinates {(1,0.333) (4, 0.333)} node [pos = 0.12, anchor = south] {$\b_1 = \b_2$};

			\addplot [thick, gray] coordinates {(1,0.333) (4, 0.333)} node [pos = 0.12, anchor = south] {$\b_1 = \b_2$};

			\addplot [PlotColor1, thick, smooth] table [col sep = comma, x index = 0, y index = 1] {signaling_1.tex} node [pos = 1, anchor = west, xshift = 0.2cm] {$b_1^{\signal}$};
			\addplot [PlotColor1, thick, smooth,forget plot] table [col sep = comma, x index = 0, y index = 1] {signaling_2.tex} node [pos = 1, anchor = west, xshift = 0.2cm] {$b_2^{\signal}$};

			\addplot [PlotColor2, thick, smooth] table [col sep = comma, x index = 0, y index = 1] {scoring_1.tex} node [anchor = west, xshift = 0.2cm] (topa) {$b_{1}^{\score}$};
			\addplot [PlotColor2, thick,smooth, forget plot] table [col sep = comma, x index = 0, y index = 1] {scoring_2.tex} node [anchor = west, xshift = 0.2cm] (bottomb) {$b_{2}^{\score}$};

			\addplot [PlotColor2, dashed,  thick, smooth] table [col sep = comma, x index = 0, y index = 1] {scoring_BR_1.tex} node [pos = 0.5, anchor = north west] (topb) {$\reg_1^{\score}$};
			\addplot [PlotColor2, dashed, thick, smooth, forget plot] table [col sep = comma, x index = 0, y index = 1] {scoring_BR_2.tex} node [pos = 0.5, anchor = south west] (bottoma) {$\reg_2^{\score}$};
	
	\end{axis}
	
	%\draw[decorate,decoration={brace, raise = 5pt}, thick, gray]	(topa) -- (topb) node [midway, anchor = west, xshift = 10pt] {$b_1$};
	%\draw[decorate,decoration={brace, raise = 5pt}, thick, gray]	(bottoma) -- (bottomb) node [midway, anchor = west, xshift = 10pt] {$b_2$};
	
	%\node[at=(plot.east), anchor=west, xshift = -0.3cm] {\pgfplotslegendfromname{leg}};
\end{tikzpicture}
\caption{Scoring---underweighting and overweighting}
\label{fig:scoring_BR}
\end{figure}

\cref{fig:scoring_BR} plots the signaling equilibrium and the scoring solution in the leading  binary example ($\s_{\e,1} = \s_{\e,2} = \s_{\d,1} = 1$) as $\s_{\d, 2}$ varies from $1$ to $2$, interpolating between the symmetric and asymmetric cases plotted in \cref{fig:obedient_b}. As distortion ability on the second feature becomes more heterogeneous, weight shifts from the second feature to first feature under both signaling and scoring. The shift is more dramatic under scoring.  \cref{fig:scoring_BR} also displays each coefficient $\reg_j^{\score} = \reg_j ( \th | \h + b^{\score} \circ \d)$, which is the weight the receiver would place on feature $j$ ex post if he could observe  the sender's features (induced by the sender's best response to the scoring rule). This plot confirms that for $\s_{\d,2} > 1$, the first feature is overweighted and the second feature is underweighted. The receiver understands this, but from the score alone he cannot disentangle the value of each feature. The exact combination of underweighting and overweighting ensures that the receiver can do no better than matching his decision with the score. 

\section{Screening} \label{sec:screening}

Having shown how the intermediary can use scoring to mitigate the receiver's commitment problem, I now study the \emph{screening} setting in which the receiver can commit to his decision rule. Screening is thus a natural benchmark, and it may also be a reasonable model of applications in which the receiver is an entrenched monopolist. I conclude this section by comparing the three settings---signaling, scoring, and screening. 

%In the scoring setting, the intermediary provides the receiver with partial commitment power by controlling his information

\subsection{Decision commitment}

The intermediary and the receiver have aligned preferences. Therefore, if the receiver has commitment power, it is optimal for the intermediary to fully disclose the sender's feature vector. The receiver's problem amounts to selecting a decision as a function of the sender's features. I restrict to linear decision rules, as in the other two settings. Recall that the intercept $b_0$ does not affect the sender's incentives, so it is optimal for the receiver to choose $b_0$ so that the mean decision error is zero. The sender's problem becomes
\begin{equation} \label{eq:screening_problem}
\begin{aligned}
	&\text{minimize}
	&& \var( b^T ( \h + b \circ \d) - \th).
\end{aligned}
\end{equation}
This screening problem is simply the scoring problem without the obedience constraint. Problem \eqref{eq:screening_problem}
has a unique solution, which I denote by $b^{\screen}$.

 \subsection{Commitment reduces distortion}
 
The contribution of the sender's distortion to the loss function in \eqref{eq:screening_problem} is $(b \circ b)^T \S_{\d\d} (b \circ b)$. Define $\| \cdot \|_{4, \d}$ on $\R^k$ by
  \[
 	\| b \|_{4, \d} =  \Bigl[ (b \circ b)^T \S_{\d\d} (b \circ b) \Bigr]^{1/4}.
 \]
As  long as $\S_{\d\d}$ has full rank, $\| \cdot \|_{4, \d}$ is a norm.\footnote{If $\S_{\d\d}$ does not have full rank, then $\| \cdot\|_{4,\d}$ is a semi-norm. The triangle inequality holds because the expression inside brackets is convex in $b$  (since $\S_{\d\d}$ is positive semidefinite and, by \ref{it:covariance},  componentwise nonnegative).} The norm $\| b \|_{4,\d}$ of the coefficient vector $b$ is one measure of the sensitivity of decisions to the sender's features. This measure captures the relevant information loss from distortion. 
%and If $\S_{\d\d}$  to show that this is in fact a norm. It measures the receiver's feature sensitivity in terms of the receiver's loss from the induced distortion. 

% It is with respect to this measure of sensitivity that the receiver's decision becomes less sensitive to the sender's features as commitment increases. 
 
\begin{thm}[Commitment reduces distortion] \label{res:reduced_distortion}
If $\S_{\d\d}$ has full rank, then
		\begin{align*}
		\| \b \|_{4,\d}
		&> \| b^{\signal} \|_{4,\d}
		\geq  \| b^{\score} \|_{4,\d}
		>  \| b^{\screen} \|_{4,\d} > 0.
		\end{align*}
The weak inequality is strict if $b^{\signal} \neq b^{\score}$. 
\end{thm}
 
As the receiver's commitment power increases, decisions become less sensitive to the sender's features, so that the sender's distortion introduces less noise into the decision.  With a single feature, \citet[Proposition 2, p.~90]{FrankelKartik2019WP} show that  the screening coefficient is smaller than the signaling coefficient.\footnote{They allow arbitrary correlation between the sender's intrinsic level (which equals quality) and distortion ability, so there may be multiple linear signaling equilibria. They show that the screening coefficient is strictly smaller than every strictly positive signaling equilibrium coefficient.} This is consistent with the inequality in \cref{res:reduced_distortion}.

\begin{comment}
sensitity of the receiver's receiver's decision becom to each feature is weights on the features are adjusted in order to make the sender's distortion less makes his decision less sensitive to the sender's features in order to deter the sender's distortion. Starting on the left side of the inequality, the regression coefficient $\b$ is the optimal coefficient vector if the sender did not distort her features at all. Next, in the signaling equilibrium, the receiver is playing a best response to the sender's distorted features, so his decision become less sensitive to the sender's features. In the scoring setting, the intermediary rearranges the weights to further reduce distortion, subject to the obedience constraint. Finally, with decision commitment, the receiver can commit to under-react to all the sender's features. 
\end{comment}

\subsection{Comparing feature weights}

\begin{figure}
		\begin{tikzpicture}
		\begin{axis}[
			clip = false,
			width=0.85 \textwidth,
			height=0.5\textwidth,
			axis y line=middle, 
			axis x line=bottom,
			clip = false,
			scale = 1.1,
			xtick = {1,4},
			xticklabels = {1,2},
			ytick = {0.333},
			yticklabels = {$\frac{1}{3}$},
			xmin = 1,
			xmax = 4.1,
			ymin = 0.245,
			ymax = 0.36,
			xlabel = {$\s_{\d,2}$},
			x label style={at={(current axis.right of origin)},anchor=west},
			y label style = {at={(current axis.above origin)},anchor=south},	
			]

				\addplot [thick] coordinates {(1,0.333) (4, 0.333)} node [pos = 0.12, anchor = south] {$\b_1 = \b_2$};

				\addplot [thick, gray] coordinates {(1,0.333) (4, 0.333)} node [pos = 0.12, anchor = south] {$\b_1 = \b_2$};

				\addplot [PlotColor1, thick, smooth] table [col sep = comma, x index = 0, y index = 1] {signaling_1.tex} node [pos = 1, anchor = west, xshift = 0.2cm] {$b_1^{\signal}$};
				\addplot [PlotColor1, thick, smooth,forget plot] table [col sep = comma, x index = 0, y index = 1] {signaling_2.tex} node [pos = 1, anchor = west, xshift = 0.2cm] {$b_2^{\signal}$};

				\addplot [PlotColor2, thick, smooth] table [col sep = comma, x index = 0, y index = 1] {scoring_1.tex} node [anchor = west, xshift = 0.2cm] (topa) {$b_{1}^{\score}$};
				\addplot [PlotColor2, thick,smooth, forget plot] table [col sep = comma, x index = 0, y index = 1] {scoring_2.tex} node [anchor = west, xshift = 0.2cm] (bottomb) {$b_{2}^{\score}$};

				\addplot [Green, thick, smooth] table [col sep = comma, x index = 0, y index = 1] {screening_1.tex} node [anchor = west, xshift = 0.2cm, yshift = -0.25cm] (topb) {$b_1^{\screen}$};
				\addplot [Green, thick, smooth, forget plot] table [col sep = comma, x index = 0, y index = 1] {screening_2.tex} node [anchor = west, xshift = 0.2cm, yshift = -0.1cm] (bottomb) {$b_2^{\screen}$};	
		
		\end{axis}
		
		%\draw[decorate,decoration={brace, raise = 5pt}, thick, gray]	(topa) -- (topb) node [midway, anchor = west, xshift = 10pt] {$b_1$};
		%\draw[decorate,decoration={brace, raise = 5pt}, thick, gray]	(bottoma) -- (bottomb) node [midway, anchor = west, xshift = 10pt] {$b_2$};
		
	\end{tikzpicture}
	\caption{Comparing commitments}
	\label{fig:comparison}
\end{figure}

I return to the setting of uncorrelated errors in order to compare scoring and screening feature-by-feature. 

\begin{thm}[Screening weights] \label{res:feature_weights}
	In the setting of uncorrelated errors, the following hold.
	\begin{enumerate}[label = (\roman*), ref = \roman*]
		\item \label{it:screening_components} The components of $b^{\screen}$ are all strictly positive, and $b_i^{\screen} > b_j^{\screen}$ if $(\s_{\e,i}, \s_{\d, i}) < (\s_{\e,j}, \s_{\d, j})$. 
		\item \label{it:screening_comparison} $b_{j}^{\screen} < b_j^{\score}$ for all $j$.
		\item\label{it:screening_weighting} $b^{\screen}$ underweights at least one feature.
		\item \label{it:screening_ratios} $w_i (b^{\screen}) \geq w_j (b^{\screen})$ if and only if $\tau_i \geq \tau_j$. 
	\end{enumerate}
\end{thm}

%		\item \label{it:score} If $k > 1$, then $b^{\score}$ and $\BR( b^{\score} )$ are incomparable. 

As under signaling and scoring, the screening coefficients are strictly positive, and the screening coefficient on a feature is higher if the intrinsic level is more accurate or distortion ability is more homogeneous (part \ref{it:screening_components}). Each screening coefficient is strictly smaller than the corresponding scoring coefficient (part \ref{it:screening_comparison}) since screening is not constrained by obedience. The ordering of the weighting ratios across features is the same under screening and scoring  (part \ref{it:screening_ratios}), but the cardinal values are different. Scoring requires both underweighting and overweighting to preserve obedience. Under screening, it is optimal to underweight at least one feature (part \ref{it:screening_weighting}), but there is no guarantee that any other feature is overweighted. 
% I wanted to show that all the weighting ratios go down together, but it doesnt work. 

\cref{fig:comparison} plots the feature weights for all three settings in the leading binary example  ($\s_{\e,1} = \s_{\e,2} = \s_{\d,1} = 1$) as $\s_{\d, 2}$ varies from $1$ to $2$. On the vertical axis, where $\s_{\d,1} = \s_{\d,2} = 1$, the signaling equilibrium and the scoring solution coincide, but the screening solution is different---it underweights both features.  If a scoring rule underweights one feature, then it must overweight the other (by obedience) and this is not optimal in the symmetric case. Moving to the right, as $\s_{\d,2}$ increases, the weight shifts from the second feature to the first. Each screening weight lies below the corresponding scoring weight. 

Finally, I consider the \emph{sender's} utility in the three commitment regimes. (Clearly, the receiver's utility increases in commitment.) The sender is risk-neutral over decisions. Under all three regimes, the expected decision is the same. Therefore, the average sender utility from a decision rule with coefficient vector $b$ is determined by the average cost of distortion
\[
(1/2) \E \sum_{j=1}^{k} (b_j \d_j)^2/\d_j
=
(1/2) (b \circ b)^T \mu_\d. 
\] 
The average sender utility is decreasing in the magnitude of each feature weight. In the setting of uncorrelated errors, we have $0 < b_j^{\screen} < b_j^{\score}$ for each $j$ (by \cref{res:feature_weights}.\ref{it:screening_comparison}), so the average sender utility is strictly higher under screening than under scoring. The reduced sensitivity lessens the sender's burden to distort her features.  While the average decision is the same under scoring and screening, decisions are more dispersed under scoring. Sender types with high distortion abilities may strictly prefer scoring to screening.

\section{Random scoring} \label{sec:extensions}

% Can random scoring help to deter distortion. There are a couple of ways to incorporate randomness. The first is to use linear scoring rules with random weights $b_0, \ldots, b_k$ random. First, the intermediary draws the weights $b_0, \ldots, b_k$. Then the intermediary applies the resulting scoring rule to sender's features. In my model, the sender is risk-neutral, so these random weights do not affect his distortion choice. Indeed, replacing any random scoring rule with its expectation will leave the sender's best response unchanged. Nevertheless, randomness can help because it affects the receiver's obedience condition. 

The main model restricts the intermediary to \emph{deterministic} linear scoring rules. Suppose instead that the intermediary can use a \emph{random} linear scoring rule, i.e., a function $ f \colon \R^k \to \D(\R)$ whose expectation is a linear function of the feature vector. That is, for some coefficients $b_0, b_1, \ldots, b_k$, we have
\[
	\E[f(x)] =  b_0 + b^T x, \qquad x \in \R^k.
\]
This class includes scoring rules with random coefficients and additive noise. The intermediary optimizes over all such rules, subject to the receiver's obedience constraint.  Since the sender is risk neutral over the decision, her best response is determined by the expectation of $f$.\footnote{If the sender were risk-averse, then using random weights could discourage distortion. \cite{EdererHoldenMeyer2018} explore such randomization in a multi-task moral hazard model.} Therefore, making $f$ random, without changing its expectation, strictly increases the mean squared decision error. It follows that the optimal random linear \emph{screening} rule must be deterministic. Scoring is more subtle because adding noise to a disobedient scoring rule could make it obedient. I show, however, that under the standing covariance assumptions the intermediary strictly prefers to achieve obedience by scaling up the coefficient vector, rather than by adding noise.

 %\footnote{By the revelation principle, this is equivalent to optimizing over all scoring rule--equilibrium pairs $(f; d,y)$ for which $y \circ f \colon \R^k \to \D(\R)$ is linear in expectation.}
 
%Relative to $f$, the scoring rule $\tilde{f}$ makes the receiver strictly worse off and leaves the sender's utility unchanged. This observation immediately proves that the optimal screening rule is deterministic, but scoring is more subtle because noise changes the receiver's obedience constraint. It may be that $\tilde{f}$ is obedient, while $f$ is not. 

%\)\var( f(\h + b \circ d) | \h + b \circ d) = 0$. 
%Let $X $ denote the induced random feature vector $\h + b \circ \d$. Scoring is noise-free  if $f(X) = \E [ f(X) | X]$, that is, the scoring rule does not add noise on-path. 
%In particular, this allows for random weights and additive noise

\begin{prop}[Random scoring] \label{res:noiseless_scoring}
	The optimal random linear scoring rule is deterministic.
\end{prop}

Therefore, the optimal scoring rule characterized in the main model remains optimal in the larger class of random linear scoring rules.

\section{Conclusion} \label{sec:conclusion}

This paper studies the design of predictive scores---like FICO credit scores---when the agents being scored behave strategically. Whenever information is dispersed and costly to acquire, it is common for intermediaries to aggregate information from different sources.  Rather than disclosing everything they learn, these intermediaries often offer simple scores. I show that with an appropriate weighting of the features, such scores can mitigate a commitment problem faced by the user of the score. In order to maximize decision accuracy, the intermediary must in general commit to use feature weights that are ex-post statistically suboptimal.

I close with two directions for future work. My model assumes a single intermediary. This is a reasonable approximation of the credit-scoring market,  but it would be interesting to study whether competing intermediaries could sustain any of the gains from scoring over signaling.  Finally, some scoring systems are kept opaque in an attempt to reduce strategic manipulation. In general,  the effect of opacity on decision accuracy is not clear. Opacity may improve accuracy if it uniformly reduces every agent's return from distortion.  On the other hand, if different agents form different beliefs about the scoring rule that is actually used, opacity could increase distortion heterogeneity and thus reduce decision accuracy.  Modeling belief-formation and optimizing behavior in response to opaque systems is an interesting challenge.

%\newpage
\appendix
%\numberwithin{equation}{section}
%\setstretch{1}

\newpage

\section{Proofs} \label{sec:proofs}

\subsection{Preliminaries}

I use the following notation. For vectors: $x \geq y$ means $x_i \geq y_i$ for all $i$; $x > y$ means $x \geq y$ and $x \neq y$; and $x \gg y$ means $x_i > y_i$ for all $i$.  For square symmetric matrices: $A \succeq B$ means $A - B$ is positive semidefinite, and $A \succ B$ means $A - B$ is positive definite. Denote the all-ones vector by $\mathbf{1}$. 

I will freely use the following implication of Assumption~\ref{it:Schur}. By the law of total variance, for any vector $b$, we have
\begin{equation} \label{x_full_rank}
\begin{aligned}
	\var( \h + b \circ \d)
	&=   \E \var(\h + b \circ \d | \d)  + \var ( \E[ \h + b \circ \d | \d]) \\
	&\succeq \E \var(\h + b \circ \d | \d)\\
	&=\E \var(\h  | \d)\\
	& \succ 0.
\end{aligned}
\end{equation}

\begin{comment}
\subsection{Proof of \texorpdfstring{\cref{res:LCE}}{Lemma \ref{res:LCE}}}

This result is implied by \cref{res:elliptical_char} (\cref{sec:elliptical_distributions}).

\subsection{Proof of \texorpdfstring{\cref{res:separating}}{Proposition \ref{res:separating}}}

This result is proved in the main text (\cref{sec:homogeneous_distortion}).

\end{comment}

\subsection{Solution concept} \label{sec:PBE}

My solution concept is \emph{linear equilibrium}, i.e., Bayes--Nash equilibrium in linear strategies. If the sender's type $(\h, \d)$ followed a full-support distribution, then every feature vector would be on-path. I assume, however, that  $\d \geq 0$ to ensure that distortion is costly for the sender. This sign restriction, combined with elliptical symmetry, forces the support of $(\h, \d)$ to be bounded. Thus, some feature vectors are off-path. My notion of linear equilibrium requires that the receiver's strategy $y \colon \R^k \to \R$ remains linear off-path. For some off-path feature vectors $x$ in $\R^k$, the receiver's decision $y(x)$ is not a best response to any belief over the type space $T = \supp (\h, \d)$.  The set of decisions that are best responses to some belief over $T$ is $Y_0 = \operatorname{conv} \big( \supp \E[ \th| \h, \d]\big)$, where $\operatorname{conv}$ denotes the convex hull. The on-path outcome of a linear signaling equilibrium can be induced by a perfect Bayesian equilibrium if and only if all sender deviations to off-path feature vectors remain unprofitable if the receiver instead takes decision $\min Y_0$ at every off-path feature vector. Computations show that this may or may not hold, depending on the distribution of $(\h, \d)$. 

On the other hand, if the model is tweaked, then every linear equilibrium can be combined with suitable beliefs to form a perfect Bayesian equilibrium. In the main model, I assume that (i) the sender does not observe her own quality $\th$, and (ii) the vector $(\th, \h, \d)$ is elliptically distributed. All the same results go through if I instead assume that (i$^\prime$) the sender \emph{does} observe her own quality $\th$, and (ii$^\prime$) the vector  $(\h, \d)$ is elliptically distributed and  $\E[\th |\h, \d]$ is a linear function of $(\h, \d)$.  It is easy to construct a distribution of $(\th, \h, \d)$ so that (ii$^\prime$) holds and $\th$ has full support. For example, let $\th$ equal a linear function of $(\h, \d)$ plus uncorrelated Gaussian noise. In this case, the agent's type is $(\th, \h, \d)$, and every decision is  best response to some belief over the type space $\supp (\th, \h, \d)$.

%.\footnote{Another approach used in the older Gaussian literature and in \cite{FrankelKartik2019WP} is to assume as a primitive that the receiver plays a linear best response even for negative values of $\d$.}

%]For example, consider the single-dimensional case, with $\E[ \th | \h] = \h$. Recall that the unique signaling equlibrium depends only on the convariance matrix $\S$. Calculations show that there is a cutoff $\k \approx 0.755$ such that if $|b^{\signal}/\b| \leq \k$, then there is a distribution covariance matrix $\S$ for which  the linear equilibrium outcome can be supported as a perfect Bayesian equilibrium. If $|b^{\signal} /\b| > \k$, there is no choice of off-path beliefs that supports the linear equilibrium outcome as a perfect Bayesian equilibrium. 

\subsection{Nonlinear equilibria} \label{sec:nonlinear_equilibria}

In the main model, I focus on linear equilibria, and I discuss why linear equilibria are appealing.  Here I construct a family of nonlinear Bayes--Nash equilibria in the single-feature setting. For some type distributions, I show that nonlinear equilibria can induce more accurate decisions than the unique linear equilibrium. The trick is that the sender's choice of distortion is informative about her intrinsic level. By contrast, in linear equilibria the sender's distortion depends only on her distortion ability.  In applications, I do not think that these particular nonlinear equilibria are realistic. The receiver punishes the sender with a low decision if the sender's feature does not exactly equal two specific values, which can be very far from the range of intrinsic levels.  Thus, these equilibria rely heavily on the stylized implicit assumptions that (a) the sender can control her feature perfectly, (b)  the receiver measures the sender's feature without error, and (c) the sender has no outside option, so she must participate. 

%Assume that $\th = \h$. 
Suppose that there is a single feature $(k = 1)$.  Let $T = \supp (\h, \d)$. Suppose that $\d$ is bounded away from $0$ over the set $T$. Fix $b > 0$. Choose $b_0$ such that $ 0 < \P (  \h + b \d \geq b_0)  < 1$. Let
\[
\D(b; b_0) = \E [ \th |\h + b \d \geq b_0] -\E [ \th |\h + b \d < b_0].
\]
For some parameters $\ubar{x}$,  $\bar{x}$, and $\ubar{y}$  to be specified below, define the receiver's strategy  $y \colon \R \to \R$  by
\[
y(x)  = \begin{cases} 
	\E [ \th | \h + b \d < b_0] &\text{if}~ x = \ubar{x},  \\
	\E [ \th | \h + b \d \geq b_0] &\text{if}~ x = \bar{x}, \\
	\ubar{y} &\text{otherwise},
\end{cases}
\]
and the sender's strategy $d \colon T \to \R$ by
\[
d ( \h, \d) =
\begin{cases}
	\ubar{x} - \h &\text{if}~  \h + b \d < b_0, \\
	\bar{x} - \h &\text{if}~ \h + b \d \geq b_0.
\end{cases}
\]
The values $\ubar{x}$ and $\bar{x}$ are given by
\begin{equation*}
	\ubar{x} =  b_0 - \frac{1}{2} \frac{\D(b; b_0)}{ b},
	\qquad
	\bar{x} =  b_0 + \frac{1}{2}\frac{\D(b;b_0)}{ b}.
\end{equation*}
The set $T$ is bounded, so the parameter $\ubar{y}$ can be chosen small enough so that no sender type finds is optimal to induce any feature value outside $\{ \ubar{x}, \bar{x}\}$. 

I claim that $(d,y)$ is a Bayes--Nash equilibrium. Clearly, $y$ is a best response to $d$. To see that $d$ is a best response to $y$, note that type $(\h, \d)$ in $T$ weakly prefers $d = \bar{x} - \h$ to $d = \ubar{x} - \h$ if and only if 
\[
\D(b, b_0) \geq \frac{( \bar{x} - \h)^2}{2 \d}  - \frac{(\ubar{x} - \h)^2}{2 \d}.
\]
Multiply both sides by $\d$  (which is strictly positive) and simplify to get
\[
\D(b, b_0)  \d  \geq \bar{x}^2/2 - \ubar{x}^2/2  - ( \bar{x} - \ubar{x}) \h,
\]
which further simplifies to $\h + b \d \geq b_0$, as needed.

For each $b > 0$, let $b_0 = \mu_\h + b \mu_\d$, and consider the associated family of  nonlinear equilibria, parameterized by $b$. As $b$ tends to $0$, the receiver's expected payoff converges to his payoff from the decision rule that equals $\E [ \th | \h \geq \mu_\h]$ if $\h \geq \mu_\h$ and equals $\E[ \th |\h < \mu_\h]$ if $\h < \mu_\h$.  For $\s_{\d}^2$ sufficiently large, the  receiver strictly prefers this decision rule to the optimal linear screening rule. It follows that for $\s_{\d}^2$ sufficiently large, there exists a nonlinear equilibrium that the receiver strictly prefers to the optimal linear screening rule, and hence also to the optimal linear scoring rule and the linear signaling equilibrium. Under the alternative specification in \cref{sec:PBE}, this nonlinear Bayes--Nash equilibrium can be supported as a perfect Bayesian equilibrium. Even under the main specification, this construction shows that for some type distributions, the receiver strictly prefers some nonlinear screening rule to her optimal linear screening rule.

%footnote{I have not constructed any nonlinear continuous equilibria.  It seems intuitive that the distorted feature vector would be more informative if the amount of distortion were increasing in $\h$, so that lower types tended to distort less and higher types tended to distort more. To generate this best response for the sender, the sensitivity of the receiver's decision rule would have to be increasing. But then the receiver's decision rule would be more sensitive to the feature vector at values that suggest greater distortion, which is not consistent with Bayesian updating.} 

\subsection{Proof of \texorpdfstring{\cref{res:existence_uniqueness}}{Theorem \ref{res:existence_uniqueness}}} \label{sec:proof_existence_uniqueness}

First I prove existence, which requires only Assumption \ref{it:Schur}. Define the chained best-response function $\BR \colon \R^k \to \R^k$ by
\[
	\BR(b) = \reg (\th | \h + b \circ \d) = \var^{-1} (\h + b \circ \d) \cov ( \h + b \circ \d, \th).
\]
By \eqref{x_full_rank}, $\var(\h + b \circ \d)$ has full rank, so $\BR$ is well-defined and continuous. It suffices to prove that $\BR$ is bounded,  for then I can restrict $\BR$ to a sufficiently large closed ball and apply Brouwer. By the orthogonality property of projection, we have
\[
	\s_\th^2 \geq \var( \BR(b) ^T (\h + b \circ \d)) = \BR(b)^T \var (\h + b \circ \d) \BR(b),
\]
for all $b$. Since $\var (\h + b \circ \d)$ is positive definite, the right side is bounded below by $\l \| \BR(b)\|^2$ for some positive $\l$. Hence $\| \BR(b)\|^2 \leq \s_\th^2 / \l$ for all $b$. 

Now I prove uniqueness. Under Assumption \ref{it:uncorrelated}, the equilibrium condition \eqref{eq:signaling_reg} can be written as
\begin{equation} \label{eq:equil_uniqueness}
\begin{aligned}
	0 &= \var(\h + b \circ \d) b  - \cov (\h + b \circ \d, \th)  \\
	&=  \S_{\h \h} b + \diag (b) \S_{\d \d} (b \circ b) - \S_{\h \th}.
\end{aligned}
\end{equation}
The right side equals $(1/2) \nabla \Phi(b)$, with the function $\Phi \colon \R^k \to \R^k$ defined by 
\begin{align*}
	\Phi(b) &= \var(b^T \h - \th) + (1/2) \var( (b \circ b)^T \d ) \\
	&= b^T \S_{\h \h} b  + (1/2) (b \circ b)^T \S_{\d\d} (b \circ b) - 2  b^T \S_{\h \th} + \s_\th^2. 
\end{align*}
I claim that $\Phi$ is strictly convex, and hence $\Phi$ has at most one stationary point. The Hessian is given by
\begin{equation} \label{eq:Hess}
	\nabla^2 \Phi(b) = 2 \S_{\h\h} + 4 \diag (b) \S_{\d\d} \diag (b) + 2 \diag ( \S_{\d\d} (b \circ b)),
\end{equation}
which is positive definite because $\S_{\h \h}$ is positive definite (by \ref{it:Schur}), and $\S_{\d\d}$ is positive semidefinite and componentwise nonnegative (by \ref{it:covariance}).

\subsection{Proof of \texorpdfstring{\cref{res:fully_informative_linear}}{Proposition \ref{res:fully_informative_linear}}}

If $(\b \circ \b)^T \d$ is constant, then there is a fully informative equilibrium with $b = \b$ and $b_0 = \b_0 - (\b \circ \b)^T \mu_\d$. 

Conversely, suppose there is a fully informative equilibrium with coefficients $(b_0, b)$. Then 
\[
b_0 + b^T ( \h + b \circ \d) = \E [ \th | \h] = \b_0 + \b^T  \h.
\]
Therefore, $(b - \b)^T \h + (b \circ b)^T \d$ is constant. Following the same steps as in \eqref{x_full_rank}, it follows that $b = \b$, and hence $(\b \circ \b)^T \d$ is constant.

\subsection{Proof of \texorpdfstring{\cref{res:info_loss}}{Proposition \ref{res:info_loss}}}

To simplify notation, set $t = \a^2$. From \eqref{eq:equil_uniqueness}, the equilibrium condition is 
\begin{equation} \label{eq:IF_equil}
0 = \S_{\h \h} b + t \diag (b) \S_{\d\d} (b \circ b) - \S_{\h \th}.
\end{equation}
Here $b$ is a function of $t$, but I suppress the dependence to simplify notation. Note that $b \neq 0$ since $\S_{\h \th} \neq 0$ (by \ref{it:correlated}). I use the implicit function theorem to compute the derivative $\dot{b}$ of $b$ with respect to $t$. Consider the right side of \eqref{eq:IF_equil}. The partial derivative with respect to $t$ is $ \diag (b) \S_{\d\d} (b \circ b)$. The partial derivative with respect to $b$ is given by
\[
D = \S_{\h \h} + 2 t \diag (b) \S_{\d\d} \diag (b) + t \diag (\S_{\d\d} (b \circ b) ).
\]
This matrix $D$ is positive definite; see the argument following \eqref{eq:Hess}. By the implicit function theorem, $b$ is a differentiable function of $t$, and the derivative is given by
\begin{equation} \label{eq:dotb}
\dot{b}  = - D^{-1} \diag (b) \S_{\d\d} (b \circ b). 
\end{equation}

The receiver's  equilibrium utility is 
\begin{align*}
	u_R 
	&= - \var ( b^T ( \h + \sqrt{t} b \circ \d) - \th) \\
	&= - b^T \S_{\h\h} b + 2 b^T \S_{\h \th} - t (b \circ b)^T \S_{\d\d} (b \circ b) -  \s_{\th}^2 .
\end{align*}
The total derivative with respect to $t$ is 
\[
	\dot{u}_R	= -(b \circ b)^T  \S_{\d \d} (b \circ b) - \Bigl[  2 \S_{\h \h} b  - 2 \S_{\h \th} +  4 t (b \circ b)^T  \S_{\d \d} \diag(b) \Bigr] \dot{b}.
\]
By \eqref{eq:IF_equil}, the term in brackets reduces to $2 t (b \circ b)^T  \S_{\d \d} \diag(b)$. Plugging in \eqref{eq:dotb}, we have
\[
	\dot{u}_R =  -(b \circ b)^T  \S_{\d \d} (b \circ b) + 2 t (b \circ b)^T \S_{\d \d} \diag(b) D^{-1} \diag(b) \S_{\d\d} (b \circ b).
\]
Fix $t > 0$. To show that $\dot{u}_R < 0$, it suffices to find some positive $\k$ such that 
\begin{equation} \label{eq:kappa}
	(1 + \k)^{-1} 2 t (b \circ b)^T \S_{\d \d} \diag(b) D^{-1} \diag(b) \S_{\d\d} (b \circ b) \leq (b \circ b)^T  \S_{\d \d} (b \circ b).
\end{equation}
The strict inequality follows because the right side is strictly positive (since $b \neq 0$ by \eqref{eq:IF_equil} and, by assumption, $\S_{\d\d}$ has full rank.)

Now I prove \eqref{eq:kappa}. Since $\S_{\h\h}$ is positive definite (by \ref{it:Schur}), we have
\[
	D \succ 2t \diag (b) \S_{\d\d} \diag (b).
\]
Choose a positive $\k$ such that 
\[
	(1 + \k)^{-1} D \succ 2t \diag (b) \S_{\d\d} \diag (b).
\]
It follows that for all $\e$ sufficiently small, we have 
\[
		(1 + \k)^{-1} D \succeq F(\e) \coloneqq 2t [\e I + \diag (b) \S_{\d\d} \diag (b) ],
\]
and hence $(1 + \k) D^{-1} \preceq F(\e)^{-1}$.  Therefore, 
\begin{align*}
	&(1 + \k) 2t  b^T  [\e I  + \diag (b) \S_{\d\d} \diag (b)]  D^{-1} [\e I + \diag(b) \S_{\d\d} \diag (b)  ] b \\
	&\leq 	2t  b^T  [\e I  + \diag (b) \S_{\d\d} \diag (b)]  F(\e)^{-1} [\e I  + \diag(b) \S_{\d\d} \diag (b)  ] b \\
	&= b^T  [\e I  + \diag (b) \S_{\d\d} \diag (b)]  b. 
\end{align*}
Passing to the limit as $\e \to 0$ yields \eqref{eq:kappa}.

\subsection{Equilibrium non-monotonicity} \label{sec:counterexamples}

\begin{exmp}[Receiver's payoff is not monotone in $\S_{\d\d}$] \label{ex:not_monotone}
	Suppose $k = 2$. Consider the covariance matrices%
	\footnote{Set $\s_{\th}^2 = 9$. This variance does not affect the analysis, as long as the induced variance matrix is positive semidefinite.}
	\[
	\S_{\th \h}
	=
	\begin{bmatrix}
		4 \\ 1
	\end{bmatrix},
	\qquad
	\S_{\h\h} 
	= 
	\begin{bmatrix}
		2 & 1 \\
		1 & 2
	\end{bmatrix}, 
	\qquad
	\S_{\d\d}
	=
	\begin{bmatrix}
		1 & 0 \\
		0 & 1
	\end{bmatrix}. 
	\]
	With these parameters, the regression coefficient $\b$ equals $(7/3, -2/3)$. Even though $\h_1$ and $\h_2$ are both positively correlated with the quality $\th$, the regression coefficient on $\h_2$ is negative because  $\h_1$ and $\h_2$ have positively correlated errors and $\h_2$ has a weaker correlation with $\th$. The signaling equilibrium is given by $b^{\signal} = (1.20,-0.10)$. As $\var(\d_2)$ increases, both components of $b^{\signal}$ shrink in magnitude, and the receiver's payoff increases, though the effect is small.%
	\footnote{Increasing $\var(\d_2)$ from $1$ to $2$ shifts $b^{\signal}$ from $(1.1951, -0.0971)$ to $(1.1950, -0.0966)$.  The receiver's utility increases from $-4.3167$ to $-4.3165$.}
\end{exmp}

\subsection{Proof of \texorpdfstring{\cref{res:scoring_equals_signaling}}{Theorem \ref{res:scoring_equals_signaling}}} \label{sec:proof_scoring_equals_signaling}

First I check that the scoring problem has a unique solution. By substituting the constraint into the objective, the scoring problem \eqref{eq:scoring_problem} can be expressed as
\begin{equation} \label{eq:scoring_proof}
	\begin{aligned}
		& \text{maximize} 
		&& b^T \S_{\h \th} \\
		& \text{subject to } 
		&&    	b^T \S_{\h\h} b + (b \circ b)^T \S_{\d\d} (b \circ b) - b^T \S_{\h \th} = 0. 
	\end{aligned}
\end{equation}
The objective is continuous and the feasible set is closed and bounded.\footnote{Suppose $b$ is feasible. Since $\S_{\d\d}$ is componentwise nonnegative (\ref{it:covariance}), we must have $b^T \S_{\h\h} b \leq b^T \S_{\h \th}$, which implies boundedness because $\S_{\h\h}$ is positive definite by \ref{it:Schur}.} Denote the constraint function by $g$. It is easy to check that $g(t \S_{\h \th})$ is strictly negative for sufficiently small, positive $t$  and $g(t \S_{\h \th})$  tends to $\infty$ as $t \to \infty$. Hence, $t \S_{\h \th}$ is feasible for some positive $t$, which means that the maximum value must be strictly positive. For uniqueness, suppose there are two solutions, denoted $b^1$ and $b^2$. Since $g$ is strictly convex, $g( (b^1 + b^2)/2) < 0$,  and hence  $t ( b^1 + b^2)/2$ is feasible for some $t > 1$, but this vector yields a strictly higher objective than $b^1$ and $b^2$, contrary to their optimality.

Now I prove \cref{res:scoring_equals_signaling}. The solution of \eqref{eq:scoring_proof} is characterized by the system\footnote{Denote the objective by $f$ and the constraint by $g$. It can be shown that $\nabla g(b) = 0$ implies $g(b) \neq 0$, so we get the necessary condition $\nabla f(b) = \l \nabla g(b)$ for some $\l$. Since $\nabla f(b) = \S_{\h \th} \neq 0$, it follows that $\l \neq 0$, so we can divide by $\l$. In the main text, I work directly with $\l^{-1}$. Since $f$ is linear, these necessary Lagrangian conditions are also sufficient.}
\begin{align}
	\label{eq:score_system_gradient} \l \S_{\h \th} - \S_{\h \h} b &= 2 \diag (b) \S_{\d\d} (b \circ b), \\
	\label{eq:score_system_constraint}  b^T \S_{\h \th} - b^T \S_{\h \h} b  &= (b \circ b)^T \S_{\d\d} (b \circ b). 
\end{align}
The signaling signaling condition \eqref{eq:equil_uniqueness} can be written as
\begin{equation} \label{eq:signal_system}
	\S_{\h \th} - \S_{\h \h }b = \diag (b)\S_{\d\d} (b \circ b).
\end{equation}	 

I prove \eqref{it:coincide} $\implies$ \eqref{it:scale} $\implies$ \eqref{it:parameter} $\implies$ \eqref{it:coincide}. Starting from \eqref{it:coincide}, $b^{\score}$ must satisfy \eqref{eq:score_system_gradient}--\eqref{eq:signal_system}. Subtract twice \eqref{eq:score_system_gradient} from \eqref{eq:signal_system} to get \eqref{it:scale}.  From \eqref{it:scale}, suppose  $b^{\signal} = t \b$ for some $t$. Substitute this expression into \eqref{eq:signal_system}. Since $\S_{\h \th} \neq 0$, we must have $t \neq 0$. Simplify \eqref{eq:signal_system} to get  \eqref{it:parameter}.  Finally, from \eqref{it:parameter}, suppose $\diag (\b) \S_{\d\d} (\b \circ \b) = t \S_{\h \th}$ for some $t$. Then \eqref{eq:score_system_gradient} and \eqref{eq:signal_system} become
\[
	(\l - 2 t)\S_{\h \th} = \S_{\h \h} b = (1 - t) \S_{\h \th},
\]
so $b^{\signal} = b^{\score} = (1 - t) \b$ and $\l = 1 + t$.

\subsection{Proof of  \texorpdfstring{\cref{res:scoring_CS}}{Proposition \ref{res:scoring_CS}}}

Consider the scoring problem in \eqref{eq:scoring_proof}. 
It suffices to  prove the strict part of the definition of decreasing in \cref{ft:decreasing} for then the weak part follows from continuity. Let $b$ denote the solution of \eqref{eq:scoring_proof}. Fix $\tilde{\S}_{\d\d} \prec \S_{\d\d}$.  Let $\tilde{g}$ and $g$ denote the respective functions on the left side of the constraint. We have $\tilde{g}(b) < g(b) = 0$. Since $\tilde{g}(t b) \to \infty$ as $t \to \infty$ and $g$ is continuous, there exists $ \hat{t} > 1$ such that $\tilde{g} ( \hat{t} b)  = 0$ and hence $\hat{t}b$ is feasible when $\var(\d) = \tilde{\S}_{\d\d}$. Finally, $(\hat{t} b)^T \S_{\h \th} > b^T \S_{\h \th} > 0$ since the optimal value of \eqref{eq:scoring_proof} is positive. 

\subsection{Proof of Theorems \ref{res:underweighting_overweighting} and \ref{res:feature_weights}}

In the setting with uncorrelated errors, the covariances are given by 
\begin{equation} \label{eq:uncorrelated_covariances}
\begin{aligned}
		\S_{\h \th} &= \mathbf{1},  \\
		\S_{\h \h} &=  \mathbf{1} \mathbf{1}^T +  \diag (\s_{\e_1}^2, \ldots, \s_{\e,k}^2), \\
		\S_{\d \d} &= \diag(\s_{\d,1}^2, \ldots, \s_{\d,k}^2).
\end{aligned}
\end{equation}
The regression, signaling, scoring, screening, and best-response vectors respectively satisfy
\begin{align}
\label{eq:reg_matrix}		\S_{ \h \th} - \S_{\h\h} \b &= 0, \\
\label{eq:signal_matrix}		\S_{\h \th} - \S_{\h \h}b &= \diag(b) \S_{\d\d} (b \circ b), \\
\label{eq:score_matrix}		\l \S_{\h \th} - \S_{\h \h}b &= 2\diag(b) \S_{\d\d} (b \circ b), \\
\label{eq:screen_matrix}		\S_{\h \th} - \S_{\h \h}b &= 2 \diag(b) \S_{\d\d} (b \circ b), \\
\label{eq:BR_matrix}		\S_{\h \th} - \S_{\h \h} \BR(b)  &= \diag(b) \S_{\d\d} \diag (b) \BR(b),
\end{align}
where in \eqref{eq:score_matrix}, $\l > 1$ by the obedience condition (since $\S_{\d\d}$ has full rank in this setting).  Plugging  in \eqref{eq:uncorrelated_covariances} gives the following componentwise system for each feature $j$:
\begin{align}
\label{eq:regression} 1 - \mathbf{1}^T \b &= \s_{\e,j}^2 \b_j,\\
\label{eq:signal_components}	1 - \mathbf{1}^T b &= \s_{\e,j}^2 b_j + \s_{\d,j}^2 b_j^3, \\
\label{eq:score_components}	\l - \mathbf{1}^T b  &= \s_{\e,j}^2 b_j + 2 \s_{\d,j}^2 b_j^3, \\
\label{eq:screen_components}	1 - \mathbf{1}^T b  &= \s_{\e,j}^2 b_j + 2 \s_{\d,j}^2 b_j^3, \\
\label{eq:BR_components}	1 - \mathbf{1}^T \BR(b) &= (\s_{\e,j}^2 + \s_{\d,j}^2 b_j^2) \BR_j (b).
\end{align}

I prove \cref{res:underweighting_overweighting} (parts \ref{it:scoring_components}--\ref{it:scoring_ratios}) and \cref{res:feature_weights} (parts \ref{it:screening_components}--\ref{it:screening_ratios}). 

For \ref{res:underweighting_overweighting}.\ref{it:scoring_components} and \ref{res:feature_weights}.\ref{it:screening_components}, consider \eqref{eq:signal_components}--\eqref{eq:screen_components}. For each equation, the right side has the same sign as $b_j$, so every component has the same sign. If $b \leq 0$ then the left side is strictly positive, giving a contradiction. Hence, $b \gg 0$. Similarly, in  \eqref{eq:BR_components}, if $b \gg 0$ then $\BR(b) \gg 0$ (since $\BR(b) \leq 0$ would give a contradiction). Finally, the comparison of features in terms of $(\s_{\e,j}, \s_{\d,j})$ follows immediately from \eqref{eq:signal_components}--\eqref{eq:screen_components}.

For \ref{res:underweighting_overweighting}.\ref{it:scoring_comparison}, apply  \cref{res:scoring_equals_signaling}. From \eqref{eq:regression}, 
$\b$ is proportional to $(\s_{\e,1}^2, \ldots, \s_{\e,k}^2)$, so condition \ref{it:parameter} in \cref{res:scoring_equals_signaling} says that $(\tau_1^2, \ldots, \tau_k^2)$ is a scalar multiple of $\mathbf{1}$. 

For \ref{res:feature_weights}.\ref{it:screening_comparison}, consider two cases. If $\mathbf{1}^T b^{\screen} < \mathbf{1}^T b^{\score}$, then $b^{\screen}_j < b^{\score}_j$ for some $j$ and hence for all $j$ by \eqref{eq:score_components}	and \eqref{eq:screen_components}.  If $\mathbf{1}^T b^{\screen} \geq \mathbf{1}^T b^{\score}$, then \eqref{eq:score_components}	and \eqref{eq:screen_components} imply $b^{\screen} \ll  b^{\score}$, which is a contradiction. 

For \ref{res:underweighting_overweighting}.\ref{it:scoring_weighting}, left-multiply  \eqref{eq:BR_matrix} by $b^T$ and subtract from the obedience condition 	\eqref{eq:score_system_constraint} to get
\[
	[b^T \S_{\h\h} + (b \circ b)^T \S_{\d\d} \diag (b)] (b - \BR(b))  = 0.
\]
For $b \gg 0$, the term in brackets is strictly positive, so the vector $b  -\BR(b)$  is either $0$ has at least one strictly positive and at least one strictly negative component. 

For \ref{res:feature_weights}.\ref{it:screening_weighting}, left-multiply \eqref{eq:screen_matrix} and \eqref{eq:BR_matrix} by $b^T$ and subtract to get
\[
[b^T \S_{\h\h} + (b \circ b)^T\S_{\d\d} \diag (b)](b - \BR(b)) = - (b \circ b)^T \S_{\d\d} (b \circ b).
\]
For $b \gg 0$, the right side is strictly negative and the vector in brackets is strictly positive, so some component of $b - \BR(b)$ must be strictly negative.

For \ref{res:feature_weights}.\ref{it:screening_ratios} and \ref{res:underweighting_overweighting}.\ref{it:scoring_ratios}, express the right side of \eqref{eq:BR_components} as $(\s_{\e,j}^2 b_j + \s_{\d,j}^2 b_j^3)/ w_j(b)$. Therefore, 
\begin{align*}
	w_i(b) \geq w_j(b)
	&\iff \s_{\e,i}^2 b_i + \s_{\d,i}^2 b_i^3 \leq \s_{\e,j}^2 b_j + \s_{\d,j}^2 b_j^3 \\
	&\iff \s_{\e,i}^2 b_i \geq \s_{\e,j}^2 b_j \\
	&\iff \tau_i \geq \tau_j.
\end{align*}
where the second and third equivalences follow from expressing the common right side of \eqref{eq:score_components} and \eqref{eq:screen_components} in two ways, as 
\[
2(\s_{\e,j}^2 b_j + \s_{\d,j}^2 b_j^3) - \s_{\e,j}^2 b_j
\quad 
\text{and}
\quad
(\s_{\e,j}^2 b_j) + 2 (\s_{\e,j}^2 b_j)^3/\tau_j^2.
\]

\subsection{Proof of \texorpdfstring{\cref{res:reduced_distortion}}{Theorem \ref{res:reduced_distortion}}}

From \eqref{eq:screen_matrix} and \ref{it:correlated}, we have
$b^{\screen} \neq 0$,  hence $\| b^{\screen} \|_{4, \d} \neq 0$ (since $\S_{\d\d}$ has full rank). First I establish the following inequalities:
\[
	\| \b \|_{4,\d}^4 
	> 
	\| b^{\signal} \|_{4,\d}^4
	\begin{cases}
	\geq \| b^{\score} \|_{4,\d}^4, \\
	> \| b^{\screen} \|_{4,\d}^4.
	\end{cases}
\]
Since $\S_{\d\d}$ has full rank, it is easy to check from \eqref{eq:reg_matrix}, \eqref{eq:signal_matrix},  and \eqref{eq:screen_matrix} that $\b \neq b^{\signal}$ and $b^{\signal} \neq b^{\screen}$. To compare $\b$ and $b^{\signal}$, recall that these vectors respectively (uniquely) minimize 
\[
	\var (b^T \h - \th) 
	\quad
	\text{and}
	\quad
	\var(b^T \h - \th) + (1/2) \| b\|_{4,\d}^4.
\]
To compare $b^{\signal}$ and $b^{\score}$, recall that these vectors respectively (uniquely) minimize
\[
\var (b^T \h - \th) +  (1/2) \| b\|_{4,\d}^4
\quad
\text{and}
\quad
\var(b^T \h - \th) +  \| b\|_{4,\d}^4,
\]
over the set of obedient scoring rules (since $b^{\signal}$ is a global minimizer over all of $\R^k$). Finally, to compare $b^{\signal}$ and $b^{\screen}$, recall that these vectors respectively (uniquely) minimize 
\[
	\var (b^T \h - \th) + (1/2) \| b\|_{4,\d}^4
	\quad
	\text{and}
	\quad
	\var(b^T \h - \th) +  \| b\|_{4,\d}^4.
\]

It remains to prove that $\| b^{\score} \|_{4,\d}^4 > \| b^{\screen} \|_{4,\d}^4$. There exists  $\l > 1$ such that $b^{\score}$ satisfies the first-order condition 
\begin{equation} \label{eq:scoring_FOC}
\S_{\h \h} b +  2 \diag (b) \S_{\d\d} (b \circ b) - \l \S_{\h \th} = 0.
\end{equation}
Viewing $\l$ as a positive parameter, \eqref{eq:scoring_FOC} implicitly defines $b$ as a function of $\l$. Taking $\l = 1$ gives the screening solution. I will show that $\| b\|_{4, \d}$ is strictly increasing in $\l$. 

The derivative of the left side of  \eqref{eq:scoring_FOC} with respect to $\l$ is $-\S_{\h \th}$. The derivative of the left side with respect to $b$ is given by
\[
	D = \S_{\h \h} + 4 \diag (b) \S_{\d\d} \diag (b) +  2 \diag (\S_{\d\d} (b \circ b)),
\]
which has full rank, as shown after \eqref{eq:Hess}. By the implicit function theorem, $b$ is locally differentiable in $\l$ and 
\[
\dot{b} = D^{-1} \S_{\h \th}.
\]

From \eqref{eq:scoring_FOC}, we have
\[
\l \S_{\h \th} - \S_{\h \h} b = 2 \diag (b) \S_{\d\d} (b \circ b). 
\]
Left multiplying by $b^T$ gives
\[
\l b^T \S_{\h \th} - b^T \S_{ \h \h} b = 2 \| b \|_{\d,4}^4.
\]
Therefore, it suffices to prove that the left side is strictly increasing in $\l$. Differentiating with respect to $\l$ gives
\[
b^T \S_{\h \th} + \l \S_{\h \th}^T \dot{b} - 2 b^T \S_{\h \h} \dot{b}.
\]
Plug in the expression for $\dot{b}$ from the implicit function theorem to get
\begin{equation} \label{eq:expr}
b^T \S_{\h \th} + \l \S_{\h \th}^T D^{-1} \S_{\h \th} - 2 b^T \S_{\h \h} D^{-1} \S_{\h \th}. 
\end{equation}

To complete the proof, I show that this expression is strictly positive. To bound the last term, I use a simple inequality for quadratic forms. If $A$ is positive semidefinite, and $x$ and $y$ are vectors, then expanding $(x - y)^T A (x - y)$ shows that $2 x^T A y \leq x^T A x + y^T A y$. Applying this here gives
\begin{align*}
2 b^T \S_{\h \h} D^{-1} \S_{\h \th} 
&= 2 \l^{-1} b^T \S_{\h \h} D^{-1} ( \l \S_{\h \th})  \\
&\leq \l^{-1} \bigl[ b^T \S_{\h \h} D^{-1} \S_{\h \h} b + (\l  \S_{\h \th})^T D^{-1} (\l \S_{\h \th}) \bigr] \\
&=  \l^{-1} b^T \S_{\h \h} D^{-1} \S_{\h \h} b + \l \S_{\h \th} D^{-1} \S_{\h \th}. 
\end{align*}
Since $D \succeq \S_{\h \h}$, the right side is bounded above by 
\[
\l^{-1} b^T \S_{\h \h} b + \l \S_{\h \th} D^{-1} \S_{\h \th}.
\]
Plugging this inequality into \eqref{eq:expr}, the last term cancels and we get the lower bound
\[
b^T \S_{\h \th} b - \l^{-1} b^T \S_{\h \h} b.
\]
Rearrange this expression and use \eqref{eq:scoring_FOC} to get
\[
\l^{-1} b^T  \bigl( \l \S_{\h \th} - \S_{\h \h} b \bigr) = \l^{-1}  (b \circ b)^T \S_{\d\d} (b \circ b).
\]
This final expression is strictly positive. 

\begin{comment}

Recall from \eqref{eq: uncorrelated_covariances} that in the setting of uncorrelated errors, the covariances are given by 
\begin{equation} \eqref{eq: uncorrelated_covariances_repeated}
\begin{aligned} 
\S_{\h \th} &= [\rho + (1 - \rho)/k] \mathbf{1} \\
\S_{\h \h} &= \rho \mathbf{1} \mathbf{1}^T + (1 - \rho)I + \diag (v_{\e}), \\
\S_{\d \d} &= \diag(v_{\d}).
\end{aligned}
\end{equation}
Plug  these into the first-order conditions to obtain
\begin{align*}
\mathbf{1} - \mathbf{1} \mathbf{1}^T  b -	\diag (\s_\e^2)b 
&= \diag(b) \diag(\s_\g^2) (b \circ b),\\
\l \mathbf{1} - \mathbf{1} \mathbf{1}^T  b  - 	\diag (\s_\e^2)b 
&= 
2 \diag(b) \diag(\s_\g^2) (b \circ b), \\
\mathbf{1} - \mathbf{1} \mathbf{1}^T  b - 	\diag (\s_\e^2)b 
&= 2 \diag(b) \diag(\s_\g^2) (b \circ b).
\end{align*}

Left-multiplying by $b^T$ shows that $\mathbf{1}^T b > 0$. Rearranging, we see that for all features $i$, we have
\begin{align*}
1  - \mathbf{1}^T  b
&= \s_{\e,i}^2 b_i + \s_{\d,i}^2 b_i^3,\\
\l  - \mathbf{1}^T  b
&= \s_{\e,i}^2 b_i + 2 \s_{\d,i}^2 b_i^3,\\
1  - \mathbf{1}^T  b
&= \s_{\e,i}^2 b_i + 2 \s_{\d,i}^2 b_i^3.
\end{align*}
\end{comment}

\subsection{Proof of \texorpdfstring{\cref{res:noiseless_scoring}}{Proposition \ref{res:noiseless_scoring}}}

Consider a random scoring rule $f$ with intercept $b_0$ and coefficient vector $b$. Let $z$ denote $\var( f(\h + b \circ \d)) - \var( b^T (\h + b \circ \d) )$. A necessary condition for obedience is 
\[
	 \var (  b^T (\h + b \circ \d)) + z = \cov (  b^T (\h + b \circ \d), \th).
\]
Otherwise, the receiver would have a linear deviation. Substituting this obedience constraint into the objective, we can express this relaxed problem as
\begin{equation}  \label{eq:relaxed_scoring}
	\begin{aligned}
		& \text{maximize} 
		&& b^T \S_{\h \th} \\
		& \text{subject to } 
		&&    	b^T \S_{\h\h} b + (b \circ b)^T \S_{\d\d} (b \circ b) + z- b^T \S_{\h \th} = 0. 
	\end{aligned}
\end{equation}
If the solution has $z > 0$, then the constraint is slack in $b$, so the first-order condition implies  $\S_{\h \th} = 0$, contrary to Assumption \ref{it:correlated}. Therefore the solution is $(b,z) = (b^{\score}, 0)$.\footnote{Technically, an optimal scoring rule could incorporate noise off-path.}

\newpage

\bibliographystyle{aer}
\bibliography{lit.bib}

@article{CitronPasquale2014,
	author = {Citron, Danielle Keats  and Frank Pasquale},
	title = {The Scored Society: Due Process for Automated Predictions},
	journal = {Washington Law Review},
	year = {2014},
	volume = {89},
	number = {1},
}

@article{Voorneveld2000,
	author = {Voorneveld, Mark},
	title = {Best-Response Potential Games},
	journal = {Economics Letters},
	year = {2000},
	volume = {66},
	number = {3},
	pages = {289--295},
}

@article{Zapechelnyuk2019WP,
	author = {Zapechelnyuk, Andriy},
	title = {Optimal Quality Certification},
	journal = {American Economic Review: Insights},
	year = {2020}, 
	volume = {2},
	number = {2},
	pages = {161--176}
}

@unpublished{Mekerishvili2018WP,
	author =  {Mekerishvili, Giorgi},
	title = {Optimal Disclosure on Crowdfunding Platforms},
	year = {2023},
	note = {Working paper}
}

@unpublished{BoleslavskyKim2018WP,
	author = {Boleslavsky, Raphael and Kyungmin Kim},
	title = {Bayesian Persuasion and Moral Hazard},
	year = {2021},
	note = {Working paper},
}

@unpublished{RodinaFarragut2016WP,
	author = {Rodina, David and John Farragut},
	title = {Inducing Effort through Grades},
	year = {2020},
	note = {Discussion Paper Series - CRC TR 224 No. 221}
}

@unpublished{Whitmeyer2019WP,
	author = {Whitmeyer, Mark},
	title = {Bayesian Elicitation},
	year = {2024},
	note = {arXiv:1902.00976}
}

@article{Perez-RichetSkreta2018WP,
	author = {Perez-Richet, Eduardo and Vasiliki Skreta},
	title = {Test Design under Falsification},
	journal = {Econometrica},
	volume = {90},
	number = {3},
	year = {2022},
	pages = {1109--1142}
}

@article{Holmstrom1999,
	author = {Holmstr\"{o}m, Bengt},
	title = {Managerial Incentive Problems: A Dynamic Perspective},
	journal = {Review of Economic Studies},
	year = {1999},
	volume = {66},
	number = {1},
	pages = {169--182},
	doi = {10.1111/1467-937X.00083}
}

@article{HornerLambertFC,
	author = {H\"{o}rner, Johannes and Nicolas Lambert},
	title = {Motivational Ratings},
	journal = {Review of Economic Studies},
	volume = {88},
	number = {4}, 
	year = {2021},
	pages = {1892--1935}
}

@unpublished{Rodina2016WP,
	author = {Rodina, David},
	title = {Information Design and Career Concerns},
	year = {2020},
	note = {Discussion Paper Series - CRC TR 224 No. 220}
}

@article{BonattiCisternasFC,
	author = {Bonatti, Alessandro and Gonzalo Cisternas},
	title = {Consumer Scores and Price Discrimination},
	journal = {Review of Economic Studies},
	volume = {87},
	number = {2},
	year = {2020}, 
	pages = {750--791}
}

@article{FischerVerrecchia2000,
	author = {Fischer, Paul E. and Robert E. Verrecchia},
	title = {Reporting Bias},
	journal = {Accounting Review},
	year = {2000},
	volume = {75},
	number = {2},
	pages = {229--245}
}

@article{BenabouTirole2006,
	author = {B\'{e}nabou, Roland and Tirole, Jean},
	title = {Incentives and Prosocial Behavior},
	journal = {American Economic Review},
	year = {2006},
	volume = {96},
	number = {5},
	pages = {1652--1678},
	DOI = {10.1257/aer.96.5.1652},
	URL = {http://www.aeaweb.org/articles?id=10.1257/aer.96.5.1652}
}

@article{QuinziiRochet1985,
	author = {Quinzii, Martine and Jean-Charles Rochet},
	title = {Multidimensional Signaling},
	year = {1985},
	journal = {Journal of Mathematical Economics},
	volume = {14},
	number = {3},
	pages = {261--284}
}

@article{Engers1987,
	author = {Engers, Maxim},
	title = {Signalling with Many Signals}, 
	journal = {Econometrica},
	year = {1987},
	volume = {55},
	number = {3},
	pages = {663--674},
	url = {http://www.jstor.org/stable/1913605},
}

@unpublished{Rick2013WP,
	author = {Rick, Armin},
	title = {The Benefits of Miscommunication in Communication Games},
	year = {2013},
	note = {Working paper},
}

@unpublished{CunninghamMorenodeBarreda2019WP,
	author = {Cunningham, Tom and Moreno de Barreda, In\'{e}s },
	title = {Effective Signal-Jamming}, 
	year = {2019},
	note = {Working paper},
}

@article{FrankelKartik2019,
	author = {Frankel, Alex and Navin Kartik},
	title = {{\bibtexorder{1}}Muddled Information},
	journal = {Journal of Political Economy},
	year = {2019},
	volume = {127},
	number = {4},
	pages = {1739--1776},
	doi = {10.1086/701604},
	URL = {https://doi.org/10.1086/701604},
}

@article{FrankelKartik2019WP,
	author = {Frankel, Alex and Navin Kartik},
	title = {{\bibtexorder{2}}Improving Information from Manipulable Data},
	journal = {Journal of the European Economic Association}, 
	volume = {20},
	number = {1}, 
	year = {2022},
	pages = {79--115}
}

@article{EdererHoldenMeyer2018,
	author =  {Ederer, Florian and Holden, Richard and Meyer, Margaret},
	title = {Gaming and Strategic Opacity in Incentive Provision},
	journal = {RAND Journal of Economics},
	year = {2018},
	volume = {49},
	number = {4},
	pages = {819--854},
	doi = {10.1111/1756-2171.12253},
	url = {https://onlinelibrary.wiley.com/doi/abs/10.1111/1756-2171.12253},
}

@article{Myerson1982,
	author = {Myerson, Roger B.},
	title = {Optimal Coordination Mechanisms in Generalized Principal-Agent Problems},
	journal = {Journal of Mathematica Economics},
	year = {1982},
	volume = {10},
	number = {1},
	pages = {67--81}
}

@article{Myerson1986,
	author = {Myerson, Roger B.},
	title = {Multistage Games with Communication},
	journal = {Econometrica},
	year = {1986},
	volume = {54},
	number = {2},
	pages = {323--358}
}

@article{Aumann1974,
	author = {Aumann, Robert J.},
	title = {Subjectivity and Correlation in Randomized Strategies},
	journal = {Journal of Mathematical Economics},
	year = {1974},
	volume = {1},
	number = {1},
	pages =  {67--96}
}

@article{Forges1986,
	author = {Forges, Fran\c{c}ois},
	title = {An Approach to Communication Equilibria},
	journal = {Econometrica},
	year = {1986},
	volume = {54},
	number = {6},
	pages = {1375--1385}
}

@article{PrendergastTopel1996,
	author = {Prendergast, Canice and Topel, Robert},
	title = {Favoritism in Organizations},
	journal = {Journal of Political Economy},
	year = {1996},
	volume = {104},
	number = {5},
	pages = {958--78}
}

@article{Cremer1995,
	author = {Jacques Cr\'{e}mer},
	title = {Arm's Length Relationships},
	journal = {Quarterly Journal of Economics},
	year = {1995},
	volume = {110},
	number = {2},
	pages = {275--295},
}

@article{DworczakDuffie2018WP,
	author = {Duffie, Darrell and Piotr Dworczak},
	title = {Robust Benchmark Design},
	journal = {Journal of Financial Economics},
	year = {2021},
	volume = {142},
	number = {2},
	pages = {775--802}
}

@inproceedings{Dalvi_etal2004,
	author = {Dalvi, Nilesh and Pedro Domingos and Mausam and Sumit Sanghai and Deepak Verma},
	title = {Adversarial Classification},
	booktitle = {Proceedings of the Tenth ACM SIGKDD International Conference on Knowledge Discovery and Data Mining},
	year = {2004},
	pages = {99--108}
}

@inproceedings{Hardt_etal2016,
	author = {Hardt, Moritz and Nimrod Megiddo and Christos Papadimitriou and Mary Wooters},
	title = {Strategic Classification},
	booktitle = {Proceedings of the 2016 ACM Conference on Innovations in Theoretical Computer Science},
	year = {2016},
	pages = {111--122},
	doi = {10.1145/2840728.2840730}
}

@inproceedings{Hu_etal2019,
	author = {Hu, Lily and Immorlica, Nicole and Vaughan, Jennifer Wortman},
	title = {The Disparate Effects of Strategic Manipulation},
	booktitle = {Proceedings of the Conference on Fairness, Accountability, and Transparency},
	year = {2019},
	pages = {259--268},
}

@article{MorrisShin2002,
	author = {Morris, Stephen and Hyun Song Shin},
	title = {Social Value of Public Information },
	journal = {American Economic Review},
	year = {2002},
	volume = {92},
	number = {5},
	pages = {1521--1534},
	DOI = {10.1257/000282802762024610},
	URL = {http://www.aeaweb.org/articles?id=10.1257/000282802762024610}
}

@unpublished{LambertMartiniOstrovsky2018WP,
	author = {Lambert, Nicolas S and Giorgio Martini and Michael Ostrovsky},
	title = {Quadratic Games},
	year = {2018},
	note = {Working paper}
}

@article{CambanisHuangSimons1981,
	author = {Cambanis, Stamatis and Steel Huang and Gordon Simons},
	title = {On the Theory of Elliptically Contoured Distributions},
	journal = {Journal of Multivariate Analysis},
	year = {1981},
	volume = {11},
	number = {3},
	pages = {368--385}
}

@book{FangKotzNg1989,
	author = {Fang, Kai-Tai and Samuel Kotz and Kai Wang Ng},
	title = {Symmetric Multivariate and Related Distributions},
	series = {Monographs on Statistics \& Applied Probability},
	volume = {36},
	year = {1989},
	publisher = {Chapman and Hall}
}

@article{OwenRabinovitch1983,
	author = {Owen, Joel and Rabinovitch, Ramon},
	title = {On the Class of Elliptical Distributions and their Applications to the Theory of Portfolio Choice},
	journal = {Journal of Finance},
	year = {1983},
	volume = {38},
	number = {3},
	pages = {745-752},
	doi = {10.1111/j.1540-6261.1983.tb02499.x},
	url = {https://onlinelibrary.wiley.com/doi/abs/10.1111/j.1540-6261.1983.tb02499.x},
}

@article{Chamberlain1983,
	author = {Chamberlain, Gary},
	title = {A Characterization of the Distributions that Imply Mean--Variance Utility Functions},
	journal = {Journal of Economic Theory},
	year = {1983},
	volume = {29},
	number = {1},
	pages = {185--201},
	doi = {10.1016/0022-0531(83)90129-1},
}

@article{Rosenthal1973,
	author = {Rosenthal, Robert W.},
	title = {A Class of Games Possessing Pure-strategy Nash Equilibra},
	journal = {International Journal of Game Theory},
	year = {1973},
	volume = {2},
	pages = {65--67},
}

@article{MondererShapley1996,
	author = {Monderer, Dov and Shapley, Lloyd S},
	title = {Potential Games},
	journal = {Games and Economics Behavior},
	year = {1996},
	volume = {14},
	number = {1},
	pages = {124--143},
	doi = {https://doi.org/10.1006/game.1996.0044},
	url = {http://www.sciencedirect.com/science/article/pii/S0899825696900445}
}

@article{BarronGoebelJensen2010,
	author = {Barron, E. N. and Goebel, R. and Jensen, R. R.},
	title = {Best Response Dynamics for Continuous Games},
	journal  = {Proceedings of the American Mathematical Society},
	year = {2010},
	volume = {138},
	number = {3},
	pages = {1069--1083},
}

@article{HofbauerSorin2006,
	author = {Hofbauer, Josef and  Sorin, Sylvain},
	title = {Best Response Dynamics for Continuous Zero-Sum Games},
	journal = {Discrete and Continuous Dynamical Systems---Series B},
	year = {2006},
	volume = {6},
	number = {1},
	pages = {215--224}
}

@unpublished{Segura-Rodriguez2021wp,
	author = {Segura-Rodriguez, Carlos},
	title = {Selling Data},
	year = {2022},
	note = {Working paper}
}

@unpublished{Bjorkegren2021wp,
	author = {Bj\"{o}rkegren, Daniel  and  Blumenstock, Joshua  E. and Knight, Samsun},
	title = {Training Machine Learning to Anticipate Manipulation},
	year = {2023},
	note = {Working paper}
}

@article{Kyle1985,
	author = {Kyle, Albert S.},
	title = {Continuous Auctions and Insider Trading},
	journal = {Econometrica},
	pages = {1315--1335},
	volume = {53},
	number = {6},
	year = {1985}
}

@incollection{Goodhart1975,
	author = {Goodhart, Charles},
	title = {Problems of Monetary Management: The U.K. Experience},
	booktitle = {Papers in Monetary Economics 1975},
	volume = {1},
	year = {1975},
	publisher = {Reserve Bank of Australia},
	address = {Sydney},
	pages = {1--20}
}

@article{Gesche2021,
title = {De-biasing Strategic Communication},
journal = {Games and Economic Behavior},
volume = {130},
pages = {452--464},
year = {2021},
author = {Tobias Gesche}
}

@article{Oyarzun2023,
	title = {Testing under Information Manipulation},
	author = {Silvia Martinez-Gorricho and Carlos Oyarzun},
	year = {2023},
	journal = {Economic Theory}
}

\newpage

\section{Online Appendix: ``Scoring Strategic Agents'' by Ian Ball} \label{sec:beyond_linear}

Nonlinear signaling equilibria are difficult to analyze in general. Here, I give a condition under which no Bayes--Nash equilibrium, pure or mixed, is fully informative.

\begin{prop}[No fully informative equilibrium] If $\S_{\d\d}$ has full rank, then the signaling game has no fully informative Bayes--Nash equilibrium. 
\end{prop}

\begin{proof} Assume $\S_{\d\d}$ is full rank. Suppose for a contradiction that the signaling game has a fully informative equilibrium. I will show that some type of the sender has a profitable deviation. 
	
	The first step is to construct the candidate deviating types. The type space $T = \supp (\h, \d)$ must contain an ellipse $E$ defined by the equation
	\[
	(\h - \mu_\h)^{T} \S_{\h\h}^{-1} (\h - \mu_\h) + (\d - \mu_\d)^T \S_{\d\d}^{-1} (\d - \mu_\d) = r^2,
	\]
	for some positive radius $r$. Choose $\h^0$ such that $(\h^0 - \mu_\h) \S_{\h \h}^{-1} ( \h^0 - \mu_\h)$ is strictly between $0$ and $r^2$. Then  $(\h^0, t \mu_\d)$  intersects $E$ for two positive values of  $t$, which I denote $t_1 < t_2$. Let $\d^0 = t_1 \mu_\d$ and set $\k = t_2/t_1$ so $\k \d^0 = t_2 \mu_\d$. Next, I construct a sequence of types converging to $(\h^0, \d^0)$ as follows. Since $\S_{\d\d}$ and $\S_{\h\h}$ both have full rank, we can find a strictly positive sequence $t^i$ converging to $0$ and a real sequence $s^i$ converging to $0$ such that each type
	\[
	(\h^i, \d^i) \coloneqq (\h^0 + t^i \b, \d^0 + s^i \d^0)
	\]
	lies on the ellipse $E$. Clearly $(\h^i, \d^i) \to (\h^0, \d^0)$ as $i \to \infty$.

	For all $i \geq 0$, choose a feature vector $x^i$ that type $(\h^i, \d^i)$ induces through some equilibrium distortion choice. Since the equilibrium is fully informative, it follows that $y(x^i) = \b_0 + \b^T \h^i$ for each $i$.  Each type $(\h^i, \d^i)$ can secure the payoff from mimicking $(\h^0, \d^0)$,  so the sequence $(x^i)$ for $i \geq 1$ is bounded. After possibly passing to a subsequence, I can assume that this sequence converges to some limit $x^\ast$. 
	
	Now I obtain the contradiction. To simplify notation, let 
	\[
	c(d) = (1/2)\sum_{j=1}^{k} d_j / \d_j^0.
	\]
	Each type $(\h^i, \d^i)$ weakly prefers $x^i$ to $x^0$, so
	\[
	t^i \| \b \|^2 \geq \frac{c( x^i - \h^i) - c( x^0 - \h^i)}{(1 + s^i)^2}.
	\]
	Passing to the limit in $i$ gives
	\begin{equation} \label{eq:x_limit}
		c(x^\ast - \h^0) \leq c(x^0 - \h^0).
	\end{equation}
	Type $(\h^0, \k \d^0)$ must be indifferent between $x^0$  and any feature vector chosen in equilibrium since $x^0$ yields same decision and cannot be more costly (for otherwise type $(\h^0, \d^0)$ would have a profitable deviation). Therefore,  type $(\h^0, \k \d^0)$ weakly prefers $x^0$ to $x^i$,  so
	\begin{equation} \label{eq:xi_expensive}
		t^i \| \b\|^2 \leq \frac{ c( x^i - \h^0) - c( x^0 - \h^0)}{\k^2} \leq \frac{ c( x^i - \h^0) - c( x^\ast - \h^0)}{\k^2},
	\end{equation}
	where the second inequality follows from \eqref{eq:x_limit}. 
	
	Similarly, since each type $(\h^i, \d^i)$ prefers $x^i$ to $x^j$, we have
	\[
	(t^i - t^j) \| \b \|^2 \geq \frac{c( x^i - \h^i) - c( x^j - \h^i)}{(1 + s^i)^2}.
	\]
	Passing to the limit as $j \to \infty$ gives 
	\begin{equation} \label{eq:xi_cheap}
		t^i \| \b \|^2 \geq \frac{c( x^i - \h^i) - c( x^\ast - \h^i)}{(1 + s^i)^2}.
	\end{equation}
	
	Clear denominators in \eqref{eq:xi_expensive} and \eqref{eq:xi_cheap} and then subtract to get
	\begin{align*}
		&((1 + s_i)^2 - (1 + \k)^2) t^i \| \b \|^2 \\
		&\geq [c( x^i - \h^i) - c(x^\ast - \h^i)] - [c(x^i - \h^0) - c (x^\ast - \h^0)] \\
		&= [c( x^i - \h^i) -  c(x^i - \h^0)] + [c( x^\ast - \h^0) - c( x^\ast - \h^i)]\\
		&=[c( x^i - \h^0 - t^i \b) -  c(x^i - \h^0)] + [c( x^\ast - \h^0) - c( x^\ast - \h^0 - t^i \b)].
	\end{align*}
	Divide by $t^i$ and pass to the limit as $i \to \infty$. By the mean value theorem, the terms on the right converge to $-c'( x^\ast - \h^0) \b$ and $c'( x^\ast - \h^0) \b$, so we obtain the contradiction 
	\[
	-(\k^2 - 1) \| \b\|^2 \geq 0.  \qedhere
	\]
\end{proof}

\end{document}